\documentclass[prb,twocolumn,showpacs,superscriptaddress,merge]{revtex4}

\usepackage[applemac]{inputenc}
\usepackage[T1]{fontenc}
\usepackage{ae}
\usepackage{graphicx} 
\usepackage{amsfonts} 
\usepackage{amssymb} 
\usepackage{amsmath} 
\usepackage{latexsym} 
\usepackage{amsthm} 
\usepackage{booktabs} 
\usepackage{psfrag} 
\usepackage{float} 
\usepackage{color} 
\usepackage{paralist} 
\usepackage{subfigure} 
\usepackage{hyperref} 
\usepackage{natbib}

\newcommand{\beq}{\begin{equation}}
\newcommand{\eeq}{\end{equation}}

\newcommand{\sign}{\,{\rm sign}}
\newcommand{\re}{\,{\rm Re}}

\newcommand{\arctanh}{\,{\rm arctanh}}

\def\bra#1{\mathinner{\langle{#1}|}} 
\def\ket#1{\mathinner{|{#1}\rangle}} 
\newcommand{\braket}[2]{\langle #1|#2\rangle}

\providecommand{\abs}[1]{\lvert#1\rvert}

\begin{document}

\title{Work distribution and edge singularities for generic time-dependent protocols in extended systems.}

\author{Pietro Smacchia}
\affiliation{SISSA, International School for Advanced Studies, via Bonomea 265, 34136 Trieste, Italy}
\affiliation{INFN, Sezione di Trieste, I-34127 Trieste, Italy}
\author{Alessandro Silva}
\affiliation{SISSA, International School for Advanced Studies, via Bonomea 265, 34136 Trieste, Italy}
\affiliation{Abdus Salam ICTP, Strada Costiera 11, 34100 Trieste, Italy}

\begin{abstract}
We study the statistics of the work done by globally changing in time with a generic protocol the mass in a free bosonic field theory with relativistic dispersion and the transverse field in the one-dimensional Ising chain both globally and locally. In the latter case we make the system start from the critical point and we describe it in the scaling limit. We provide exact formulas in all these cases for the full statistics of the work and we show that the low energy part of the distribution of the work displays an edge singularity whose exponent does not depend on the specifics of the protocol that is chosen, and may only depend on the position of the initial and final value with respect to the critical point of the system. We also show that the condensation transition found in the bosonic system for sudden quenches [A. Gambassi and A. Silva, Phys. Rev. Lett. {\bf 109}, 250602 (2012)] is robust with respect to the choice of the protocol.
\end{abstract}

\pacs{05.70.Ln,05.30-d}

\maketitle

\section{Introduction}

The interest in out-of-equilibrium coherent dynamics of quantum many-body systems has grown impressively in recent years, making this field one of the most active in quantum physics. Recent attention to this problem has been mainly triggered by recent experimental advances, especially in the context of cold-atoms physics \cite{bloch_review}. These are systems very weakly coupled to the environment, in such a way that decoherence effects are highly suppressed. One of the first experiments in this context, the observation of  the collapse and revival of a system driven across the Mott-superfluid transition \cite{greiner2002b}, provided a strong indication of the high degree of coherence retained by the system even in the presence of strong interactions. In turn, the nature of these interactions has been conjectured to have important consequences on the possible thermalization of such isolated systems, as suggested by the observation of a lack of thermalization in a one-dimensional bosonic gas \cite{kinoshita}.  The 
closeness to integrability is in particular behind the phenomenon of pre-thermalization \cite{Kollar2011,*Berges2004,*Gring2012,*Kitagawa2011}.

The non-equilibrium dynamics is expected to depend on the way the system is taken out of equilibrium. For a thermally isolated system the most natural procedure is to vary in time a parameter $\lambda$ of the Hamiltonian. In this setting there is still a large amount of freedom in the choice of the way that parameter is changed from its initial to its final value, i.e., in the choice of the specific protocol. The two extreme cases are an instantaneous change, the so-called {\it quantum quench} and an adiabatic protocol, and they are the most studied cases in the literature, while more generic time-dependent protocols are hardly addressed. However, their study is important to understand what dynamical features are generic or if there are changes in the behavior of the system when different protocols are considered (for example, some features can depend on how fast the parameter is varied in time). Generic protocols can also be important for applications in quantum information \cite{harlander_2011, cardy_2011} and quantum optimization problems \cite{caneva_2011}, where one looks for the best protocol in order to achieve a certain goal, which usually is put in the form of maximizing a certain figure of merit. Finally, their study can be useful in dealing with concrete experimental situations in which a sudden variation of the parameters may be difficult to achieve realistically.


A second important point to keep in mind is that system parameters can be varied either globally or locally. In both cases this variation causes the emission of quasiparticles traveling across the system and causing the spread of correlations with a certain velocity (whose maximum value is given by the Lieb-Robinson bound \cite{lieb_robinson}). The difference between global and local variations  is that in the former case this emission happens everywhere in space and the system is excited by an extensive amount of energy, while in the latter scenario the emission is restricted to the point or region where the quench is performed, which behaves like a ``quantum antenna" \cite{harlander_2011}. Qualitatively, the ballistic propagation of signals results in the so-called ``light-cone" effect \cite{calabrese_06}. With this picture in mind it is clear that the effects of a quench are particularly strong at a critical point of the system, where excitations are gapless.

The effects of nonequilibrium protocols have been characterized in a variety of ways, for example by looking at the evolution of correlations functions \cite{calabrese_06} or at entanglement entropies \cite{calabrese_05, *calabrese_07, *cincio_07,*stephan_2011}. From a fundamental point of view a quantum quench is, however, nothing but a thermodynamic transformation, and can therefore be characterized by the work done on the system \cite{Silva2008, Gambassi2011, GS12, Spyros}, the entropy produced \cite{Polkovnikov2011a}, and the heat that may have been exchanged \cite{Bunin2011}. In the following, in order to characterize the energy spectrum of the excitations created, we will focus on the work, which for a non equilibrium protocol is a stochastic quantity fluctuating among different realizations of the same protocol, and so is described by a probability distribution $P(W)$ \cite{Campisi2011,*Jarzynski1997,*Kurchan}. 

Moreover, the statistics of the work is widely recognized as a valuable tool to study universal behavior~\cite{Bunin2011} and/or detect dynamical phase transitions \cite{Heyl2012b}.  In particular, it has been shown that for sudden quenches ending near a critical point of a system, it is connected to the so-called critical Casimir effect, implying that the low-energy part of $P(W)$ displays universal features \cite{GS12, Gambassi2011,Spyros}. Moreover, for systems of bosons such a universal behavior is possible even for large values of the work due to the appearance of a condensation transition \cite{GS12} analogous to Bose-Einstein condensation at equilibrium. In this context a natural question that arises is how robust these features are with respect to different choices of the protocol (i.e., different variations of the system parameters in time). 

In this work we address in detail this question computing the statistics of the work for generic protocols and for global and local variations of the system parameters 
in integrable systems, which can be represented in terms of free fermions (or Majorana fermions) or free bosons. This class of models describes systems of experimental interests, including interacting bosons and fermions in one dimension \cite{Cazalilla2011} and relative phase fluctuations in split condensates \cite{gritsev_07,*Gritsev_07a}. One of the main universal features emerging for abrupt quantum quenches, a power law edge singularity characterizing the low-energy part of $P(W)$, is shown to be hardly sensitive to the details of the protocol considered, being characterized by an exponent that depends only on the initial and final values of the parameter being varied. Moreover we show that the above-mentioned condensation transition is robust with respect to the choice of the protocol.

The rest of the paper is organized as follows. In Sec. \ref{sec:general} we discuss the definition and the general properties of the probability distribution of the work done $P(W)$. In Sec. \ref{sec:bosons} we compute the statistics of the work in a free bosonic field theory with a relativistic dispersion and a time-dependent mass. In Sec. \ref{sec:ising_global} we consider the same quantity for a one-dimensional quantum Ising chain with a time-dependent transverse field. In Sec. \ref{sec:local} we consider the same system subjected now to a local time-dependent change of the transverse field and briefly discuss the possibility of extending our findings to more complex models. In the case of the local quench, in addition to the statistics of the work, we also consider the transverse magnetization and its correlations produced by the quench. The main results of Sec. \ref{sec:local} have already been published in \onlinecite{Smacchia2012}. Section \ref{sec:conclusions} summarizes and discusses the results.

\section{Generalities on the statistics of the work}
\label{sec:general}

In order to set the notation, let us consider a system described by the Hamiltonian $H[\lambda(t)]$, where $\lambda$ is a time-dependent parameter that has an initial value $\lambda_0$ at time $t=0$ and a final value $\lambda_1$ at final time $t=\tau$. We will keep the function $\lambda(t)$, i.e. the nonequilibrium protocol, unspecified, though we will assume the initial state to be the ground state $\ket{0}_0$ of the initial Hamiltonian. Also, the parameter can be  changed either globally or locally. The probability distribution of the work  $W$ done on the system is in general given by~\cite{Talkner2007,Campisi2011}
\beq
P(W)= \sum_n \delta\left(W-E_n(\tau)+E_0(0) \right) \lvert {}_\tau\! \langle n| U(\tau) \ket{0}_0 \rvert^2,
\label{eq:pw}
\eeq
where $\ket{n}_t$ are the instantaneous eigenstates of $H[\lambda(t)]$ with eigenvalues $E_n(t)$ and $U(\tau)$ is the evolution operator from $t=0$ to $t=\tau$.  As it is evident from this definition, $P(W)$ has a threshold value given by $E_0(\tau)-E_0(0)$, so in the following we will refer $W$ to this value in such a way that $W \geq 0$.

Introducing the moment generating function $G(s)$,
\beq
G(s)= \langle e^{-s W} \rangle ,
\eeq
which exists $\forall s \geq\bar{s}$, where $\bar{s} \leq 0$ is a parameter to be determined case by case.  We have that 
\beq
G(s)=\bra{\psi(\tau)}e^{-s \tilde{H}[\lambda(\tau)]}\ket{\psi(\tau)},
\label{eq:char_func}
\eeq
where $\tilde{H}[\lambda(\tau)]=H[\lambda(\tau)]-E_0(\tau)$ in such a way that the ground state has zero energy, and $\ket{\psi(\tau)}=U(\tau)\ket{0}_0$ is the evolution at time $\tau$ of the initial state. 

Following Refs. \onlinecite{GS12} and \onlinecite{Gambassi2011},  we can use the quantum to classical correspondence to interpret the function $G(s)$ for $s>0$ as a partition function in a $(d+1)$-dimensional slab of thickness $s$ of a classical system, with transfer matrix $e^{-\tilde{H}[\lambda(\tau)]}$ and equal boundary conditions described by $|\psi(\tau) \rangle$. The cumulant generating function $F(s)=\ln G(s)$ assumes, up to a minus sign,  the role of  free energy and in the case of global protocols it is useful to consider the free energy density per unit area $f(s)=-L^{-d} F(s)$, which can be decomposed in decreasing power of $s$ as,
\beq
f(s)=2 f_s+f_c(s).
\label{eq:f(s)}
\eeq
Here the bulk contribution (proportional to $s$) is absent (we rescaled the variable $W$ to have a threshold value equal to zero), while $f_s$ is the surface free energy associated with the two identical boundaries and $f_c$ is the Casimir effect contribution, which represents an effective interaction between the two boundaries that goes to zero for large $s$. 

From the definitions above we can already derive a few general features of the probability distribution $P(W)$.  The latter will have a peak at the origin whose weight is $P_0=e^{-2 L^d f_s}=\lvert \braket{\psi(\tau)}{0}_\tau \rvert^2$, which is just the fidelity of the evolved state and represents the probability to end up in the ground state of the final Hamiltonian. In the following we will frequently discuss a quantity connected to the fidelity, the {\it normalized logarithmic fidelity} per unit volume 
\beq
\hat{f}_s=\ln \lvert \langle \psi(\tau) |0\rangle_{\tau}  \rvert L^{-d} \frac{(2 \pi)^d}{\Omega_d},
\label{eq:log_fidelity}
\eeq
where $\Omega_d$ is the solid angle in $d$ dimensions.

In addition to this feature at the origin, one expects an edge singularity at $W=\Delta$, where $\Delta$ is the minimum energy gap of the final Hamiltonian. The behavior close to this threshold will be determined by the behavior of $f_c$ for large $s$.  For abrupt quenches this power-law edge singularity turns out to be universal in the sense of statistical mechanics~\cite{GS12,Gambassi2011}. One of the main results of this paper will be to show that this edge singularity is also ``universal" in the time domain, i.e., independent of the specific details of the protocol considered. These edge singularities are, however, relevant only in finite-size systems, while their spectral weight is exponentially small in the system volume. 

As the system size increases, it is in turn convenient to consider the statistics of the work density $w=W/L^d$. 
The key quantity to study in this context is the rate function $I(w)$, 
whose importance  lies in the fact that for $L \rightarrow \infty$ we have $p(w) \sim \exp\left[-L^d I(w)\right]$ \cite{Touchette2009}.
Since for large $L$ we can perform the inverse Laplace transform via a saddle-point approximation, the rate function turns out to be given by the Legendre-Fenchel transform of $f(s)$,
\beq
I(w)=-{\rm inf}_s [s w-f(s)],
\eeq
where the infimum is taken in the domain of definition of $G(s)$. Therefore, in this case the universality observed in the edge singularities for the statistics of the work close to the lowest threshold is inherited by the universal large-deviation statistics of $I(w)$  below the average work done $\overline{w}$. 
In addition, in the case of bosonic systems a universal behavior is also observed for large deviations above the average work done. In this case a transition associated with a $p(w)$ changing character from exponential to algebraic decay,  i.e., $I(w > \overline{w}) \rightarrow 0$, has been predicted for a system of free bosons and for sudden quenches, as the starting point of the quench tends to the critical point.  Since this behavior has been shown to be analogous to Bose-Einstein condensation in the grand-canonical ensemble \cite{GS12}, we will call this phenomenon {\it condensation transition}. In the following we will show its robustness with respect to the choice of the protocol.

The case of a local protocol differs substantially from the global one because, as already stated in the Introduction, the work done on the system is not extensive. Local quenches are particularly interesting in critical systems with  gapless excitations, where even a local change of the Hamiltonian can have important effects. 
In particular, in the case of a local protocol in a gapless system, if one excludes the special case of cyclic protocols, $P(W)$ will not generally have a $\delta$ peak at the origin in the thermodynamic limit, because of the Anderson orthogonality catastrophe \cite{anderson}, and the low-energy part is expected to consist in an edge singularity starting right at $W=0$, whose form will still be determined by the large $s$ behavior of $\ln G(s)$. 
 
In the case of the quantum Ising model we show that this edge singularity is independent of the specifics of the protocol and we provide an argument to generalize this result to other systems. We stress that differently from the case of global protocols, the low-energy part of the statistics of the work can retain a considerable spectral weight also in the thermodynamic limit, due to the non extensive amount of energy injected in the system. 
 
\section{Free Bosonic Theory}
\label{sec:bosons}

In the following we will study the robustness of the features described, which were originally discussed in the context of abrupt quenches, to a modification of the protocol. In order to do so, let us first start with the case of a generic protocol in a free bosonic system. In this section we consider  a free bosonic Hamiltonian diagonalizable in independent momentum modes as,
\beq
H_B[m(t)] = \frac{1}{2} \int \frac{d^d k}{(2 \pi)^d} \left[\pi_k^2+\omega^2_k(t) \phi_k^2 \right],
\eeq
where the integral runs over the first Brillouin $\abs{k}<\pi$, $[\phi_k,\pi_{k'}]=i \delta_{k,k'}$, and we assume a relativistic dispersion relation $\omega_k(t)=\sqrt{k^2+m^2(t)}$. This simple model captures the physics of a number of physical
systems, ranging from ideal harmonic chains to the low-energy properties of interacting fermions and bosons in
one dimension \cite{Cazalilla2011} and split condensates \cite{gritsev_07,*Gritsev_07a} . We will consider generic protocols that take the mass from an initial value $m_0$ to a final value $m_1$ in a time $\tau$. The case of a sudden quench has been analyzed in detail in Ref. \onlinecite{Spyros}.

Since the different $k$ modes are decoupled, we have that $G(s)=\prod_k G_k(s)$, where $G_k(s)$ represents the moment generating function of a single mode, which is that of a single quantum harmonic oscillator with time-dependent frequency. Therefore, we will now compute the moment generating function in this simple system. Though for a simple harmonic oscillator the same problem has already been considered in Ref. \onlinecite{Lutz1008}, here we derive the results for a single mode using a different method that will be applicable also to the case of fermions.

\subsection{Moment generating function for an harmonic oscillator}
\label{sec:single_o}

Let us start by considering a single quantum harmonic oscillator with time-dependent frequency
\beq
H_o(t)=\frac{1}{2} p^2+\frac{1}{2} \omega^2(t) x^2,
\label{eq:ho}
\eeq
where $\omega(0)=\omega_0$, $\omega(\tau)=\omega_1$, and $x$ and $p$ are the usual position and momentum operators satisfying $[x,p]=i$.

At each time $t$ the Hamiltonian is diagonal in terms of the well-known bosonic operators
\beq
\begin{aligned} a_t &=& \sqrt{\frac{\omega(t)}{2}} \left(x+\frac{i}{\omega(t)} p\right), \\
a^\dagger_t & = & \sqrt{\frac{\omega(t)}{2}}  \left(x-\frac{i}{\omega(t)} p \right),
\end{aligned}
\eeq
obeying the commutation relation $[a_t,a^\dagger_t]=1$, which allow us to write the Hamiltonian as
\beq
H_o(t)=\omega(t) \left(a^\dagger_t a_t+\frac{1}{2}\right).
\label{eq:ho_a}
\eeq
The initial state is assumed to be the ground state $\ket{0}_0$ of $H_o(0)$, such that $a_0 \ket{0}_0=0$. 

In order to compute the moment generating function $G(s)$ using Eq. (\ref{eq:char_func}) we have to write the state $\ket{\psi(\tau)}$ in terms of the operators $a^\dagger_\tau$ and $a_\tau$ that diagonalize the final Hamiltonian. For this purpose we will introduce a time-dependent operator $\tilde{a}(t)$ that annihilates the state $\ket{\psi(t)}=U(t) \ket{0}_0$, i.e.,
\beq
\tilde{a}(t)\ket{\psi(t)}=0. 
\label{eq:tilde_cond}
\eeq
This operator exists since, for quadratic Hamiltonians, Gaussian states retain their nature during the evolution. Moreover, this operator is characterized by being constant in the Heisenberg representation. Indeed, taking a time derivative of Eq. (\ref{eq:tilde_cond}), one can prove that $i \frac{\partial}{\partial t} \tilde{a}(t)\ket{\psi(t)}=-[\tilde{a}(t),H_o(t)]\ket{\psi(t)}$, which implies that, in the subspace spanned by $\ket{\psi(t)}$,
\beq
i\frac{d}{d t} \tilde{a}^H(t)=0,
\label{eq:t_ind}
\eeq
where $\tilde{a}^H(t)=U^\dagger(t) \tilde{a}(t) U(t)$ is the operator $\tilde{a}(t)$ in Heisenberg representation (the superscript $H$ will be used in the following always to indicate the Heisenberg evolution of an operator).
  
Since we know how $\tilde{a}(\tau)$ acts on $\ket{\psi(\tau)}$, the computation of $G(s)$ proceeds by finding the relation between the operators $a_\tau$ and $a^\dagger_\tau$, diagonalizing the Hamiltonian at the final time $\tau$, and the operators $\tilde{a}(t)$ and $\tilde{a}^\dagger(t)$, whose action on the evolved state is easy. 

In order to do so, we consider the equations of motion for $a^H_\tau(t)$ and $a^{\dagger,H}_\tau(t)$.  Using the single boson Bogoliubov transformation
\beq
\begin{split}
 a_t=&\frac{1}{2} \left(\sqrt{\frac{\omega_1}{\omega(t)}}+\sqrt{\frac{\omega(t)}{\omega_1}} \right) a_\tau\\
 &-\frac{1}{2} 
\left(\sqrt{\frac{\omega_1}{\omega(t)}}-\sqrt{\frac{\omega(t)}{\omega_1}} \right) a^\dagger_\tau, 
\end{split}
\label{eq:bogoliubov}
\eeq
we can rewrite the Hamiltonian as
\beq
H_o(t)=\frac{\omega_1^2+\omega^2(t)}{2 \omega_1} a^\dagger_\tau a_\tau+\frac{\omega^2(t)-\omega^2_1}{4 \omega_1} \left(a^2_\tau+{a^\dagger_\tau}^2\right)+{\rm const}.
\eeq
From this expression one can easily compute the commutator $[a_\tau,H_o(t)]$ and obtain the evolution equation
\beq
 i\frac{d}{dt} a^H_\tau(t) = \frac{\omega_1^2+\omega(t)^2}{2 \omega_1} a^H_\tau(t)+\frac{\omega^2(t)-\omega_1^2}{2 \omega_1}[a^\dagger_\tau(t)]^H.
\label{eq:a_evolution}
\eeq
To solve this equation let us make the ansatz
\beq
a_\tau^H(t)=\alpha(t)\tilde{a}^H(t)+\beta^\star(t) \tilde{a}^{\dagger,H}(t).
\label{eq:a_ansatz}
\eeq
Putting Eq. (\ref{eq:a_ansatz}) into Eq. (\ref{eq:a_evolution}) and considering Eq. (\ref{eq:t_ind}), we finally find the evolution equation for the coefficients $\alpha(t)$ and $\beta(t)$,
\begin{subequations}
\beq
i\frac{d}{d t} \alpha(t)=\frac{\omega_1^2+\omega^2(t)}{2 \omega_1} \alpha(t)+\frac{\omega^2(t)-\omega_1^2}{2 \omega_1} \beta(t),
\eeq
\beq
i\frac{d}{d t} \beta(t)=-\frac{\omega_1^2+\omega^2(t)}{2 \omega_1} \beta(t)+\frac{\omega^2_1-\omega^2(t)}{2 \omega_1} \alpha(t).
\eeq
\label{eq:alpha_beta_ev}
\end{subequations}
The initial conditions are given by the coefficients of the Bogoliubov transformation connecting the operators diagonalizing the Hamiltonian (\ref{eq:ho}) at the final time $t=\tau$ with the operators $\tilde{a}_0$ and $\tilde{a}_0^\dagger$. However, since we are assuming to start with the initial ground state, we have $\tilde{a}(0)=a_0$, so that the initial conditions can be read from Eq. (\ref{eq:bogoliubov}), getting
\beq
\alpha(0)=\frac{1}{2} \left(\sqrt{\frac{\omega_1}{\omega_0}}+\sqrt{\frac{\omega_0}{\omega_1}} \right), \; \beta(0)=\frac{1}{2} \left(\sqrt{\frac{\omega_1}{\omega_0}}-\sqrt{\frac{\omega_0}{\omega_1}} \right).
\label{eq:initial_bos}
\eeq
The last step for computing the moment generating function $G(s)$ is to write the evolved state $\ket{\psi(\tau)}$ in terms of the operators $a_\tau$ and $a^\dagger_\tau$. Since this state is annihilated by $\tilde{a}(\tau)$, which is related to $a_\tau$ and $a^\dagger_\tau$ by the translation of Eq. (\ref{eq:a_ansatz}) at time $\tau$ into the Schroedinger picture, that is,
\beq
a_\tau=\alpha(\tau)\tilde{a}(\tau)+\beta^\star(\tau) \tilde{a}^{\dagger}(\tau),
\eeq
it must be quadratic in terms of $a_\tau$. One indeed finds
\beq
\ket{\psi(\tau)}=\frac{1}{\sqrt{\abs{\alpha(\tau)}}} \exp \left(\frac{\beta^\star(\tau)}{2 \alpha^\star(\tau)} (a^\dagger_\tau)^2 \right) \ket{0}_\tau,
\label{eq:boson_psi_evolved}
\eeq
where $a_\tau \ket{0}_\tau=0$. 

Since we have now expressed the evolved state in terms of the operators that diagonalize the final Hamiltonian, we can readily compute $G(s)$ from Eq. (\ref{eq:char_func}) (for example, using coherent states generated by $a^\dagger_\tau$ \cite{Spyros}), getting 
\beq
G(s)=\frac{1}{\abs{\alpha(\tau)}\sqrt{1-\abs{\lambda(\tau)}^2 e^{-2 s \omega_1}}},
\label{eq:g_single_o}
\eeq
with $\lambda(\tau)=\frac{\beta(\tau)}{\alpha(\tau)}$, which is defined for $s \geq \frac{\ln \abs{\lambda(\tau)}}{\omega_1}$.

\subsection{Moment generating function for the bosonic theory}

\begin{figure}
\begin{center}
\includegraphics[width=\columnwidth]{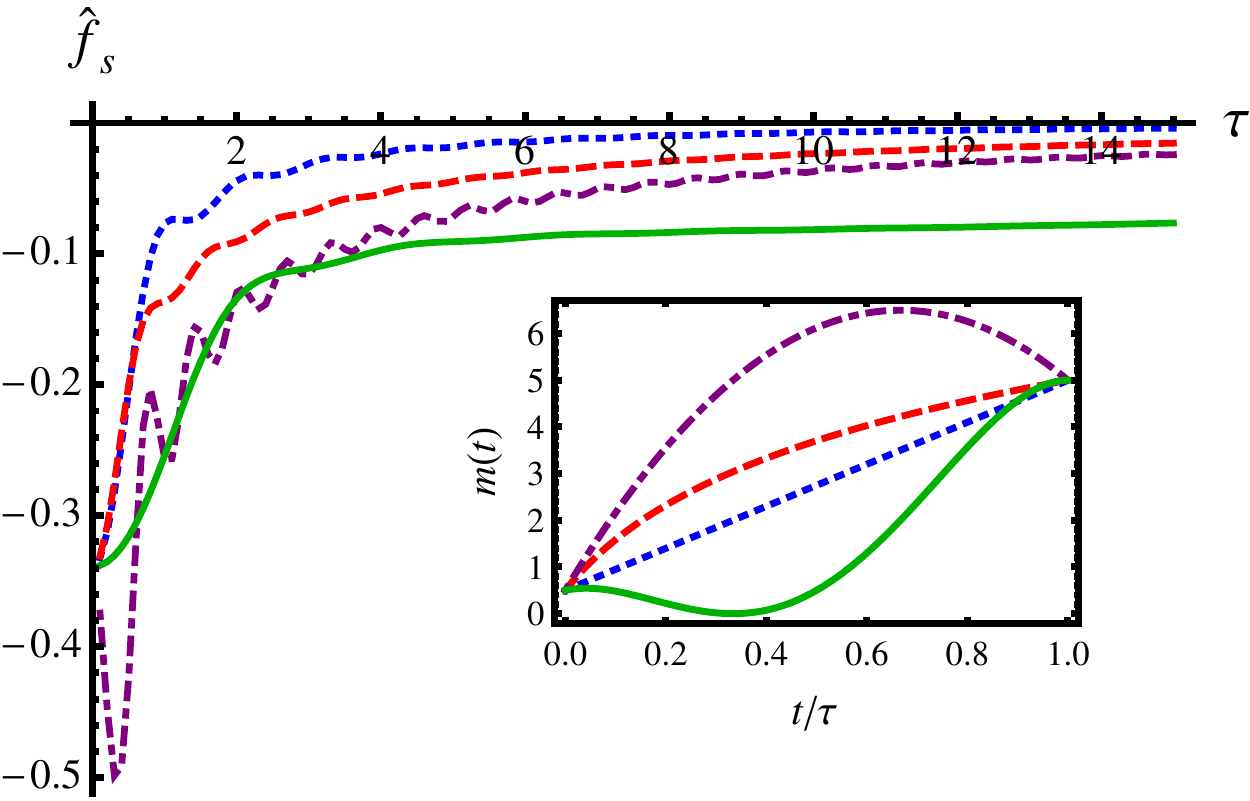}
\caption{(Color online) Plot of $\hat{f}_s$ [see Eq. (\ref{eq:log_fidelity})],for different protocols as a function of the duration $\tau$, with $m_0=0.5$ and $m_1=5$. The considered protocols are defined in Eq. (\ref{eq:protocols}) and shown in the inset. In particular the dotted (blue) one is $m_{\rm lin}$, the dashed (red) one is $m_{\rm log}$, the  dotted-dashed (purple) one is $m_{\rm par}$ and the solid (green) one is $m_{\rm quart}$}
\label{fig:fidelity}
\end{center}
\end{figure}

Using the result of the previous section, it is now easy to write down the full cumulant generating function, which is given by
\beq
\frac{\ln G(s)}{L^d}=-\frac{1}{2} \int \frac{d^d k}{(2 \pi)^d} \ln \left[\frac{1-\abs{\lambda_k(\tau)}^2 e^{-2 s \omega_k(\tau)}}{1-\abs{\lambda_k(\tau)}^2 }\right],
\label{eq:f_bosons}
\eeq
where $\lambda_k$ is defined in the previous section for each mode $k$ and the function is defined for $s>\bar{s}_B={\rm sup}_k \frac{\ln\abs{\lambda_k(\tau)}}{\omega_k(\tau)}$. Following Sec. \ref{sec:general}, we can identify the two contribution $f_c(s)=\frac{1}{2} \int_k \ln[1-\abs{\lambda_k(\tau)}^2] e^{-2 s \omega_k(\tau)}$ and $f_s=-\frac{1}{2} f_c(0)$.

We observe that for an adiabatic protocol, since the final state is the ground state of the final Hamiltonian, we would have $\lambda_k(\tau)=0$ $\forall k$, so that the function $P(W)$ would simply become a $\delta$ function at the origin as expected. For a sudden quench, since the state does not change and remains in the initial ground state, we would have $\lambda_k(\tau)=\lambda_k(0)$, whose actual value can be read from Eq. (\ref{eq:initial_bos}) for each mode $k$ and is in agreement with previous results \cite{Spyros, GS12}.

For a generic protocol, using Eqs. (\ref{eq:alpha_beta_ev}), one can find an evolution equation of Riccati type that fully determines the function $\lambda_k(\tau)$ and so the full distribution function,
\beq
\begin{split}
i \frac{d}{dt} \lambda_k(t)=&-\frac{\omega^2_k(\tau)+\omega^2_k(t)}{\omega_k(\tau)} \lambda_k(t)\\
&+\frac{\omega^2_k(\tau)-\omega^2_k(t)}{2 \omega_k(\tau)} \left[1+\lambda_k^2(t) \right],
\label{eq:lambda_ev}
\end{split}
\eeq
with initial condition $ \lambda_k(0)=\frac{\beta_k(0)}{\alpha_k(0)}$.

 When $m_1 \rightarrow 0$ we have that $\omega_k(\tau) \rightarrow k$, so the coefficients of Eq. (\ref{eq:lambda_ev}) become divergent for $k \rightarrow 0$. For this reason it is convenient to make the substitution
\beq
x_k(t)=\frac{1}{\omega_k(\tau)} \frac{1+\lambda_k(t)}{1-\lambda_k(t)},
\label{eq:lambda_x}
\eeq
with the new variable satisfying the elegant Riccati-like equation
\beq
i \frac{d}{dt} x_k(t)=-\omega^2_k(t) x_k^2(t)+1,
\label{eq:x_ev}
\eeq
with an initial condition, $x_k(0)=1/\omega_k(0)$, fully determined by the initial parameters.

Let us now compare different protocols considering first the normalized log-fidelity defined in Eq. (\ref{eq:log_fidelity}).  We will consider a linear, a logarithmic, a parabolic, and a quartic protocol, given by
\begin{subequations}
\beq
m_{\rm lin}(t)= m_0+(m_1-m_0) \frac{t}{\tau},
\eeq
\beq
m_{\rm log}(t)= m_0+(m_1-m_0) \frac{\ln (1+6 t/\tau)}{\ln 7},
\eeq
\beq
m_{\rm par}(t)=m_0+(m_1-m_0) \left(4 \frac{t}{\tau}-3 \frac{t^2}{\tau^2}\right),
\eeq
\beq
m_{\rm quart}(t)=m_0 +(m_1-m_0) \sum_{n=1}^4 \rho_n (t/\tau)^n,
\eeq
\label{eq:protocols}
\end{subequations}
with $\rho_n$ in the last protocol chosen in such a way that the function has a minimum with zero mass at $t/\tau=1/3$. The actual values of these constant can be found in Appendix \ref{sec:coefficients}, while the various protocols in the case of $m=0.5$ and $m_1=5$ are plotted in the inset of Fig. \ref{fig:fidelity}

In Fig. \ref{fig:fidelity} the log-fidelity is shown for different protocols as a function of the total duration $\tau$, taking $m_0=0.1$ and $m_1=5$. From this figure we see that for the linear and logarithmic protocols the log-fidelity is essentially an increasing function of $\tau$ tending to zero (implying a fidelity tending to one); for logarithmic protocols it is always lower than for a linear one. In the parabolic case we see oscillations for small $\tau$, when it is possible to have a fidelity lower than in the sudden case, while in the quartic case the fidelity decreases quite rapidly at the beginning, but then reaches a plateau at a value different from zero. This is due to the fact that, since this protocol touches the critical point $m=0$, where the system is gapless, it is not possible to have adiabatic behavior.

\begin{figure*}
\begin{center}
\subfigure[\label{fig:work_bosons}]{\includegraphics[width=\columnwidth]{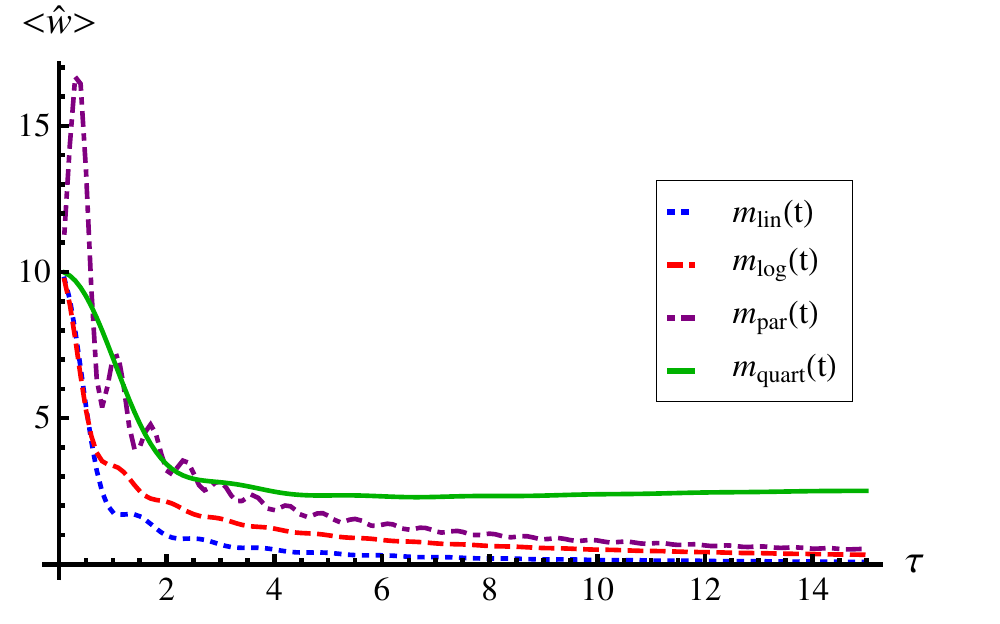}}
\subfigure[\label{fig:var_bosons}]{\includegraphics[width=\columnwidth]{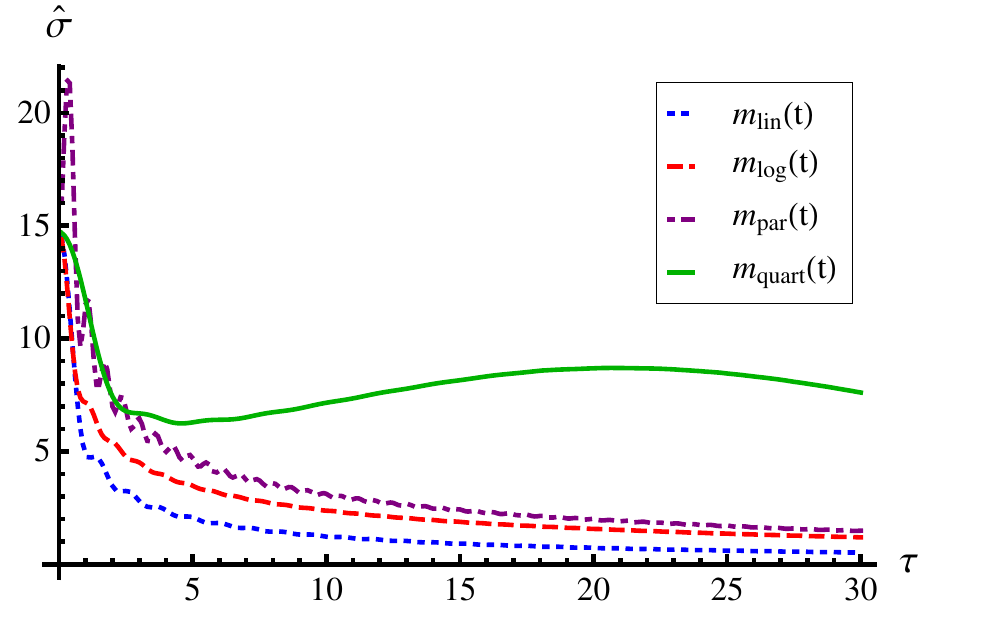}}
\caption{(Color online) Plot of (a) $\langle \hat{w} \rangle$ and (b) $\hat{\sigma}^2$  [see the definition below Eq. (\ref{eq:cumulants_boson})] for the different protocols defined in Eqs. \ref{eq:protocols} as a function of the duration $\tau$, with $m_0=0.5$ and $m_1=5$.}
\label{fig:cumulants}
\end{center}
\end{figure*}

Let us now consider the cumulants of the distribution $P(W)$. Using the formula
\beq
k_n=(-1)^n \frac{\partial^n}{\partial s^n} \ln G(s)_{\rvert_{s=0}},
\label{eq:cumulants}
\eeq
with $k_n$ representing the $n$th cumulant, and Eq. (\ref{eq:bos_form}) one obtains
\begin{subequations}
\beq
k_1=\langle W \rangle= L^d \int \frac{d^d k}{(2 \pi)^d} \frac{\abs{\lambda_k(\tau)}^2  \omega_k(\tau)}{1-\abs{\lambda_k(\tau)}^2} ,
\eeq
\beq
k_2=\sigma^2= L^d \int \frac{d^d k}{(2 \pi)^d} \frac{2 \abs{\lambda_k(\tau)}^2 \omega_k^2(\tau)}{[1-\abs{\lambda_k(\tau)}^2]^2},
\eeq
\beq
\frac{\sigma}{\langle W \rangle} \sim L^{-d/2},
\eeq
\beq
k_3=L^d \int \frac{d^d k}{(2 \pi)^d} 4 \frac{ \omega_k^3(\tau)[\abs{\lambda_k(\tau)}^4+\abs{\lambda_k(\tau)}^2]}{\left[1-\abs{\lambda_k(\tau)}^2\right]^3},
\eeq
\beq
\frac{k_3}{\sigma^3} \sim L^{-d/2}.
\eeq
\label{eq:cumulants_boson}
\end{subequations}
We notice that all the cumulants are extensive, i.e., proportional to the volume $L^d$, which is a consequence of the function $\ln G(s)$ itself being extensive. For this reason it is more appropriate to study the probability distribution of the work per unit volume \cite{GS12} $w=W/L^d$ , which has as a moment generating function $\tilde{G}(s)=G(s/L^d)$, so that the cumulants $\tilde{k}_n$ of this intensive variable are given by $\tilde{k}_n=L^{d(1-n)} k_n$. Therefore, in the limit of large $L$ the probability distribution of $w$ will become a Gaussian function with average value $k_1$ and variance $k_2/L^d$ that tends to zero as $L\rightarrow \infty$.

In Fig. \ref{fig:cumulants} we plot the first two cumulants per unit volume normalized by a geometric factor, i.e., $\langle \hat{w} \rangle=L^{-d} \langle w \rangle \frac{(2 \pi)^d}{\Omega_d}$ and $\hat{\sigma}^2 =L^{-d} \sigma \frac{(2 \pi)^d}{\Omega_d}$ for the different protocols defined in (\ref{eq:protocols}), taking $m_0=0.5$ and $m_1=5$. We see that the qualitative behavior of the two cumulants is the same: in the case of the linear and logarithmic protocols they are essentially decreasing functions of $\tau$ that tend to zero for $\tau$ large and with the logarithmic cumulants always bigger than the linear ones. This is expected since the larger $\tau$ is the more adiabatic the protocol is and the less work is done on the system; in the case of the parabolic protocol there are oscillations for small $\tau$ that rapidly decrease in amplitude so that the cumulants are larger than the sudden case only for small duration. We notice also that the value of the cumulants for the parabolic protocol is 
always larger than the linear and logarithmic ones. Finally, the cumulants for the quartic protocol at the beginning decrease quite fast; then in the case of the average there is essentially a plateau that seems to slightly decrease for large values of $\tau$, while for the variance the plateau is replaced by an increase of the function. The last protocol, except for small $\tau$, always has larger values of both the cumulants. The different qualitative behavior of the quartic protocol has again to be ascribed to the impossibility of achieving an adiabatic behavior.

Another interesting feature to be considered is the asymptotic behavior of $P(W)$ for small $W$ that, as explained above, is expected to display an edge singularity, which is determined by the asymptotics of $f(s)$ for large $s$. Apart from the constant $2 f_s$, this is just the asymptotic behavior of $f_c(s)$. In particular we will now study how the edge singularity is affected by the choice of a specific protocol.

We start by expanding the logarithm as $\ln \left[1-\abs{\lambda_k(\tau)}^2 e^{-2 s \omega_k(\tau)} \right]=-\sum_{n=1}^\infty e^{-2 s n \omega_k(\tau)} \abs{\lambda_k(\tau)}^2/n$. Then, since $\abs{\lambda_k(\tau)}^2 \leq 1$, we can interchange the order of the integration and the sum because of the convergence of the series. For $m_1\neq 0$ we have
\beq
\begin{split}
& f_c(s)=\frac{1}{2} \sum_{n=1}^\infty \int_k e^{-2 s n \omega_k(\tau)} \frac{\abs{\lambda_k(\tau)}^{2 n}}{n}\\
&  \simeq  \frac{1}{2} \sum_{n=1}^\infty e^{-2 s n m_1} \frac{\abs{\lambda_0(\tau)}^{2 n}}{n} \left(\frac{m_1}{4 \pi s n}\right)^{d/2},
\end{split}
\eeq 
where the integrals has been evaluated in the stationary phase approximation. The full series can be written as (${\rm Li}$ denotes the polylogarithm or Jonquiere's function)
\beq
f_c(s) \simeq \frac{1}{2} \left(\frac{m_1}{4 \pi s}\right)^{d/2} Li_{1+d/2} \left[e^{-2 s m_1} \abs{\lambda_0(\tau)}^2 \right],
\eeq
while the leading asymptotic behavior is given by the first term
\beq
f_c(s) \simeq \frac{e^{-2 s m_1}}{2} \left(\frac{m_1}{4 \pi s} \right)^{d/2} \abs{\lambda_0(\tau)}^2.
\eeq
From this we can extract the form of the edge singularity at the threshold. Indeed, we have that
\beq
\begin{split}
G(s) \simeq \, & e^{-2 L^d f_s} \Biggl[1+\\
& L^d \, \frac{e^{-2 s m_1}}{2} \left(\frac{m_1}{4 \pi s} \right)^{d/2} \abs{\lambda_0(\tau)}^2\Biggr],
\end{split}
\eeq
implying
\beq
\begin{split}
P(W)= &e^{-2 L^d f_s}\Biggl[\delta(W)+ \\
& L^d \left(\frac{m_1}{4 \pi}\right)^{d/2} \frac{ \abs{\lambda_0(\tau)}^2}{2 \Gamma(d/2)} \frac{\Theta(W-2 m_1)}{(W-2 m_1)^{1-d/2}} +\dots \Biggr].
\end{split}
\eeq
The most interesting feature is that the exponent of the edge singularity is completely determined by the dimensionality, independently of the choice of the protocol, which only affects the coefficient through the absolute value of $\lambda_0(\tau)$. Moreover, as we will show in more details in the next section, in the case of a protocol starting from the critical point $m_0=0$ we have that $\abs{\lambda_0(\tau)}^2=1$ independently of the details of the protocol, so in this case also the coefficient (apart from the overall factor) of the edge singularity is not affected by the choice of the protocol. We also observe that the edge singularity becomes milder and milder as the dimensionality $d$ of the system is increased, turning from a divergence for $d<2$ to a vanishing distribution for $d>2$.

\begin{figure}
\begin{center}
\includegraphics[width=\columnwidth]{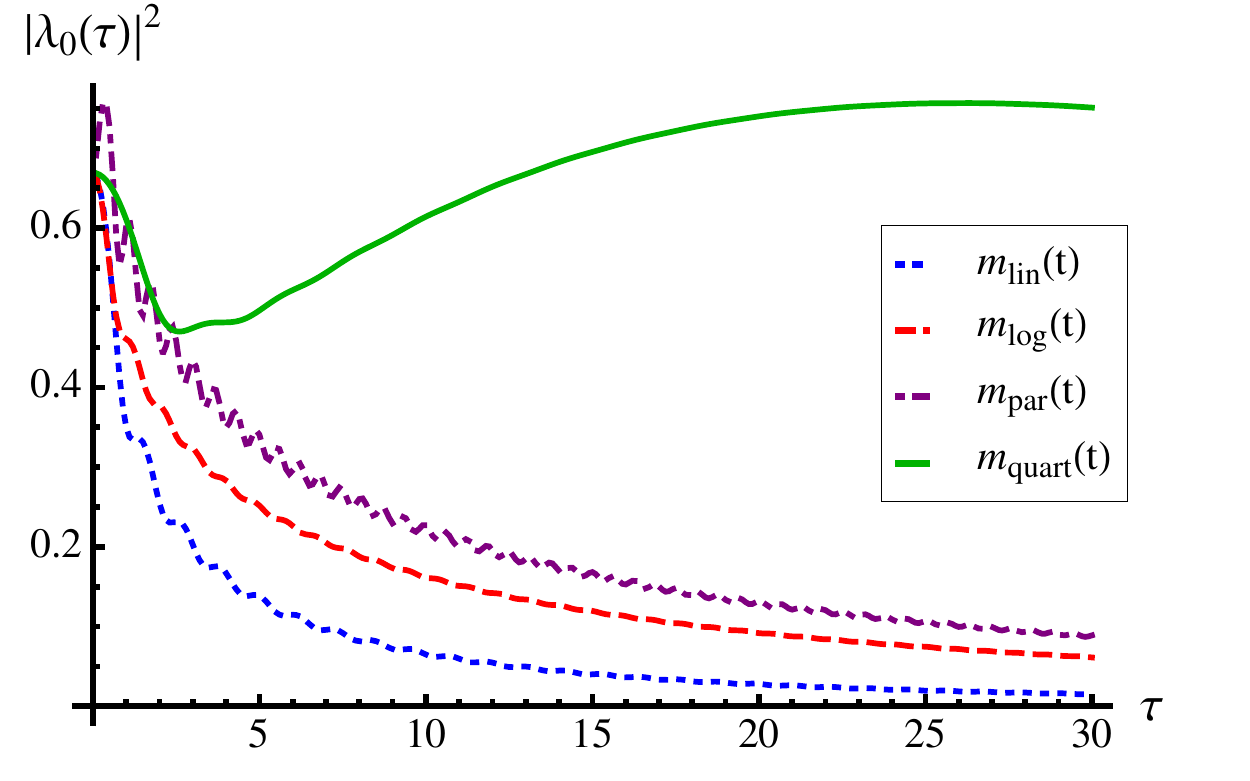}
\caption{Plot of the coefficient of the edge singularity $\abs{\lambda_0(\tau)}^2$ for the different protocols defined in Eqs. (\ref{eq:protocols}) with $m_0=0.5$ and $m_1=5$, as a function of $\tau$.}
\label{fig:edge}
\end{center}
\end{figure}

In Fig. \ref{fig:edge} we plot the value of $\abs{\lambda_0(\tau)}^2$ for the protocols defined by Eqs. (\ref{eq:protocols}) as a function of $\tau$. We see that for the linear, logarithmic, and parabolic protocols it decreases to zero, with the latter showing oscillations for small $\tau$; in the case of the quartic protocol, after an initial decrease, it increases and seems to reach a plateau. This is again a consequence of the fact that the protocol touches the critical point $m=0$, where the mode $0$ is gapless.

In the case of zero final mass, i.e., $m_1=0$, we have
\beq
\begin{split}
f_c(s)=& \frac{1}{2} \sum_{n=1}^\infty \int_k e^{-2 s n \abs{k}} \frac{\abs{\lambda_k(\tau)}^{2 n}}{n}\\
&  \simeq  \frac{\Omega_d}{(2\pi)^d} \frac{1}{2} \sum_{n=1}^\infty \frac{\Gamma(d)}{(2 s n)^d},
\end{split}
\eeq 
where we used $\abs{\lambda_0(\tau)}^2=1$, which is a simple consequence of Eq. (\ref{eq:lambda_x}). The full series is now given by
\beq
f_c(s) \simeq \frac{\Omega_d}{(2\pi)^d} \frac{\Gamma(d)}{(2 s)^d} \zeta(d),
\eeq
with leading asymptotic behavior
\beq
f_c(s) \simeq \frac{\Omega_d}{(2\pi)^d} \frac{1}{2} \frac{\Gamma(d)}{(2 s)^d},
\eeq
which, similarly to the previous case, gives for the distribution of the work the result
\beq
\begin{split}
P(W)= &e^{-2 L^d f_s}\Bigl[\delta(W)+ \frac{\Omega_d}{(2\pi)^d} L^d \frac{1}{2^{d+1} W^{d-1}}+\dots \Bigr].
\end{split}
\eeq
Thus, in this case the edge singularity is exactly at the origin, as expected from the final Hamiltonian being gapless, and both the exponent and the coefficient (apart from the overall factor) are independent of the choice of the protocol.

\subsection{Condensation Transition}
\label{sec:condensation}

In this section we study the robustness of the condensation transition discussed in Sec. {\ref{sec:general} , showing that such a transition is present for every protocol starting at the critical point $m_0=0$.

As already discussed in Sec. \ref{sec:general}, the key quantity to study is the rate function
\beq
I(w)=-{\rm inf}_s \left[sw -f(s)\right],
\eeq
where the infimum is taken in the domain of definition of $f(s)$, $s>\bar{s}_B$, and $s_B$ is a non positive number defined below Eq. (\ref{eq:f_bosons}), 
\beq
\bar{s}_B={\rm sup}_k \frac{\ln\abs{\lambda_k(\tau)}}{\omega_k(\tau)}.
\label{eq:s_bar}
\eeq
The rate function $I(w)$ is a concave function with a minimum for $w=\langle w \rangle$ and asymptotic behavior for $w \gg \langle w \rangle$  determined by $\bar{s}_B$, i.e., $I(w) \simeq \bar{s}_B w$. As we can see from Eq. (\ref{eq:initial_bos}), $m_0=0$ implies $\lambda^2_k(0) = 1 + O(k)$; therefore, for a sudden quench $\bar{s}_B=0$, implying that $I(w)=0$ for $w> \langle w \rangle$ (this is finite for $d > 1$). The vanishing of $I(w)$ means that the decay of $p(w)$ becomes algebraic and as a result the cumulants with $n \geq d$ diverge \cite{GS12}.

Asking whether the transition is still present for a generic protocol is equivalent to asking if $\bar{s}_B$ is still zero, which, as can be read from Eq.(\ref{eq:s_bar}), is equivalent to saying that $\abs{\lambda_0(\tau)}=1$. 

In order to address this question we write $\lambda_0(t)= \rho(t) e^{i \theta(t)}$ and use Eq. (\ref{eq:y_ev}) for $k=0$ to derive the equations for the modulus and the phase. We get
\begin{subequations}
\beq
\frac{d}{dt} \rho(t)=-\frac{m^2(t)-m_1^2}{2 m_1} \sin \theta(t) \left(\rho^2(t)-1\right)
\eeq
\beq
\frac{d}{dt} \theta(t) = \frac{m_1^2+m^2(t)}{m_1}+\frac{m^2(t)-m_1^2}{2 m_1} \cos \theta(t) \left(\frac{1}{\rho(t)}+\rho(t) \right).
\eeq
\label{eq:initial_critical_bosons}
\end{subequations}
 We clearly see that $\rho=1$ is a stationary solution. Therefore, for every protocol with $m_0=0$, since the initial condition is $\rho(0)=1$ we have that $\lambda_0(\tau)=1$ and the transition is still present.

\section{Global protocols in the Ising chain}
\label{sec:ising_global}

Let us now show that the existence of edge singularities and their independence from the details of the protocol pertain also to global protocols in a one-dimensional quantum Ising chain, described by the Hamiltonian 
\beq
H_I [g(t)]=-\frac{1}{2} \sum_i \left[ \sigma_i^x \sigma^x_{i+1}+g(t) \sigma_i^z \right],
\label{eq:ham_ising}
\eeq
where $\sigma_i^{x,z}$ represent the Pauli matrices and the time-dependent transverse field $g(t)$ is changed from an initial values $g_0$ to a final value $g_1$. This model is a prototypical, exactly solvable example of a quantum phase transition, whose critical point is $g_c=1$, separating a quantum paramagnetic phase ($g>1$) from a quantum ferromagnetic phase ($g<1$) characterized by a non vanishing values of the order parameter $\langle \sigma^x \rangle$ \cite{Sachdev_book}. 

Performing a Jordan-Wigner transformation $\sigma_i^+=\prod_{j<i} (1-2 c^\dagger_j c_j) c^\dagger_i$, with $\sigma_i^+=(\sigma_i^x+i \sigma_i^y)/2$, followed by a Fourier transform, one can write the Hamiltonian (\ref{eq:ham_ising}) as
\beq
H_I[g(t)]= \sum_{k>0} \begin{pmatrix} c^\dagger_k& c_{-k} \end{pmatrix} \tilde{H}_k(t) \begin{pmatrix} c_k \\ c^\dagger_{-k} \end{pmatrix},
\eeq
where $c_k$ and $c^\dagger_k$ are fermionic operators obeying the usual commutation relations $\{c_k,c^\dagger_{k'} \}= \delta_{k,k'}$ and $\{c_k,c_{k'}\}=0$, and the matrix $\tilde{H}_k$ is given by 
\beq
\tilde{H}_k(t)= \begin{pmatrix} g(t)-\cos k & -\sin k\\ -\sin k & \cos k-g(t) \end{pmatrix}.
\eeq
As in Sec. \ref{sec:bosons}, the model is reduced to a non interacting one. Hence, as before, we may first focus on the computation of the moment generating function $G(s)$ for a single mode $k$.

\subsection{Moment generating function for a single fermionic mode}
\label{sec:single_ising}

The first step in the computation of the moment generating function for a single fermionic mode $G_k(s)$ is to find the equivalent of the operators $a_t$ and $a^\dagger_t$, and so to find the operators that diagonalize the Hamiltonian at each time $t$, which we will call $\gamma^t_k$ and $(\gamma^t_k)^\dagger$. These are connected to the Jordan-Wigner fermions by the well-known Bogoliubov transformation
\beq
\begin{pmatrix}
c_k \\ c^\dagger_{-k}
\end{pmatrix}= \begin{pmatrix} u_k(t) & -v_k(t)\\
v_k(t) & u_k(t) \end{pmatrix} 
\begin{pmatrix}
 \gamma^t_k \\ (\gamma^{t}_{-k})^\dagger
\end{pmatrix},
\eeq
where $u^2_k(t)+v^2_k(t)=1$. 

The coefficients of the transformation have to be chosen in such a way that $\begin{pmatrix} u_k(t) & v_k(t) \end{pmatrix}^T$ and $\begin{pmatrix} -v_k(t) & u_k(t) \end{pmatrix}^T$ are the eigenvectors of $\tilde{H}_k$ with eigenvalues $\epsilon_k(t)$ and $-\epsilon_k(t)$ respectively, where $\epsilon_k(t)=\sqrt{1+g^2(t)-2 g(t) \cos k}$. Therefore, we have
\begin{subequations}
\beq
u_k(t)=\frac{1}{\sqrt{2}} \sqrt{1+\frac{g(t)-\cos k}{\epsilon_k(t)}},
\eeq
\beq
v_k(t)=-\frac{1}{\sqrt{2}} \sqrt{1-\frac{g(t)-\cos k}{\epsilon_k(t)}}.
\eeq
\end{subequations}
After this transformation the Hamiltonian for the single mode becomes
\beq
H_k(t)= \epsilon_k(t) \left[(\gamma^t_k)^\dagger \gamma^t_k+(\gamma^t_{-k})^\dagger \gamma^t_{-k} -1 \right).
\label{eq:h_single_ising}
\eeq
The computation now proceeds through essentially the same steps done in Sec. \ref{sec:single_o} translated into a fermionic language \cite{dziarmaga_05}.

Thus, with the goal of finding the expression of the evolved state $\ket{\psi(t)}$ in terms of the operators diagonalizing the final Hamiltonian, we start by defining the operators $\tilde{\gamma}^t_k$ and $\tilde{\gamma}^t_{-k}$ as the ones that annihilate the evolved state at time $t$, i.e.,
\beq
\tilde{\gamma}_{\pm k}(t) \ket{\psi(t)}=0,
\label{eq:tilde_condition_ising}
\eeq
where $|\psi(t)\rangle=U(t) \ket{0}_0$ and $\ket{0}_0$ is the initial ground state satisfying the condition $\gamma^0_{\pm k} \ket{0}_0=0$. As in the case of bosons, the condition (\ref{eq:tilde_condition_ising}) implies $i \frac{d}{dt} \tilde{\gamma}^{t,H}_{\pm k}(t)=0$ in the subspace spanned by $\ket{\psi(t)}$.

As in the case of bosons, we now look for the connection between the $\tilde{\gamma}_k(t)$ and $\gamma^\tau_k$'s. In order to do so we first look for the equation of motion for the Heisenberg  operators $\gamma^{\tau,H}_k(t)$ and $[\gamma^{H}_{-k}(t)]^\dagger$, obtained by computing the commutators of these operators with the Hamiltonian $H_k[g(t)]$. Using the Bogoliubov transformation
\beq
\begin{pmatrix} \gamma_k^t \\ (\gamma_{-k}^t)^\dagger \end{pmatrix}=
\begin{pmatrix}
 \mu_k(t) & \nu_k(t) \\
-\nu_k(t) & \mu_k(t) 
\end{pmatrix}
\begin{pmatrix} \gamma_k^\tau \\ 
(\gamma^\tau_{-k})^\dagger
\end{pmatrix}
\label{eq:bogoliubov_fermion}
\eeq 
with
\begin{subequations}
 \beq
\mu_k(t)=u_k(\tau) u_k(t)+v_k(\tau) v_k(t),
\eeq
\beq
\nu_k(t)=u_k(\tau) v_k(t)-v_k(\tau) u_k(t),
\eeq
\label{eq:mu_nu_k}
\end{subequations}
we find
\beq
\begin{split}
H_k(t)= \begin{pmatrix} (\gamma^\tau_k)^\dagger & \gamma^\tau_{-k} \end{pmatrix} \begin{pmatrix} r_k(t) &  s_k(t) \\ s_k(t) & -r_k(t) \end{pmatrix} \begin{pmatrix} \gamma_k^\tau \\ (\gamma^\tau_{-k})^\dagger \end{pmatrix},
\end{split}
\eeq
with
\begin{subequations}
 \beq
r_k(t)=\frac{g(t) g_1-\cos k [g(t)+g_1]+1}{\epsilon_k(\tau)},
\eeq
\beq
s_k(t)=\frac{\sin k\left[g(t)-g_1\right]}{\epsilon_k(\tau)}.
\eeq
\label{eq:r_s}
\end{subequations}
One can now easily derive the equations of motion
\beq
i\frac{d}{dt}
\begin{pmatrix}
\gamma^{\tau,H}_k(t)\\
[\gamma^{\tau,H}_{-k}(t)]^\dagger
\end{pmatrix}
= \begin{pmatrix}
   r_k(t) & s_k(t)\\
 s_k(t) & -r_k(t)
  \end{pmatrix}
\begin{pmatrix}
 \gamma_k^{\tau,H}(t) \\
(\gamma^{\tau,H}_{-k}(t))^\dagger
\end{pmatrix}.
\label{eq:gamma_ev}
\eeq
Using the ansatz
\beq
\begin{pmatrix}
 \gamma^{\tau,H}_k(t)\\
[\gamma^{\tau,H}_{-k}(t)]^\dagger
\end{pmatrix}
=\begin{pmatrix}
  a_k(t) & -b^\star_k(t)\\
b_k(t) & a^\star_k(t) 
 \end{pmatrix}
\begin{pmatrix}
 \tilde{\gamma}_k^H(t) \\
\tilde{\gamma}_{-k}^{\dagger,H}(t)
\end{pmatrix},
\label{eq:gamma_gammatilde}
\eeq 
 we finally find the equations for the coefficients,
\begin{subequations}
\beq
i\frac{d}{dt} a_k(t)=r_k(t) a_k(t)+s_k(t) b_k(t),
\eeq
\beq
i\frac{d}{dt} b_k(t)=-r_k(t) b_k(t)+s_k(t) a_k(t),
\eeq
\label{eq:a_b_ev}
\end{subequations}
where the initial conditions are given by the coefficients of the Bogoliubov transformation that connects the operators diagonalizing the Hamiltonian (\ref{eq:h_single_ising}) at the final time and the operators annihilating the initial state. Since we made the hypothesis of starting from the ground state of the initial Hamiltonian, the latter operators are those that diagonalize the Hamiltonian at time $t=0$ and so we get from Eq. (\ref{eq:bogoliubov_fermion})
\beq
a_k(0)=\mu_k(0), \quad b_k(0)=\nu_k(0).
\eeq
The last step is to express the evolved state $\ket{\psi(\tau)}$ in terms of the operators $\gamma^\tau_{\pm k}$, using the relation between these operators and the $\tilde{\gamma}^\tau_{\pm k}$, which is given by the Scroedinger version of Eq. (\ref{eq:gamma_gammatilde}). In this way one finds 
\beq
\ket{\psi(\tau)}=\left[a^\star_k(\tau)+b^\star_k(\tau) (\gamma^\tau_{-k})^\dagger (\gamma^\tau_k)^\dagger \right] \ket{0}_\tau,
\eeq
where $\gamma^\tau_{\pm k} \ket{0}_\tau=0$.

Now that we know the evolved state in terms of the operators that diagonalize the final Hamiltonian, the computation of $G(s)$ from Eq. (\ref{eq:char_func}) can be straightforwardly done, obtaining
\beq
G(s)=\abs{a_k(\tau)}^2 \left(1+\abs{y_k(\tau)}^2 e^{-2 s \epsilon_k(\tau)} \right),
\eeq
with $y_k(\tau)=\frac{b_k(\tau)}{a_k(\tau)}$.
\\

\subsection{Moment generating function for the Ising chain}

\begin{figure}
\begin{center}
\includegraphics[width=\columnwidth]{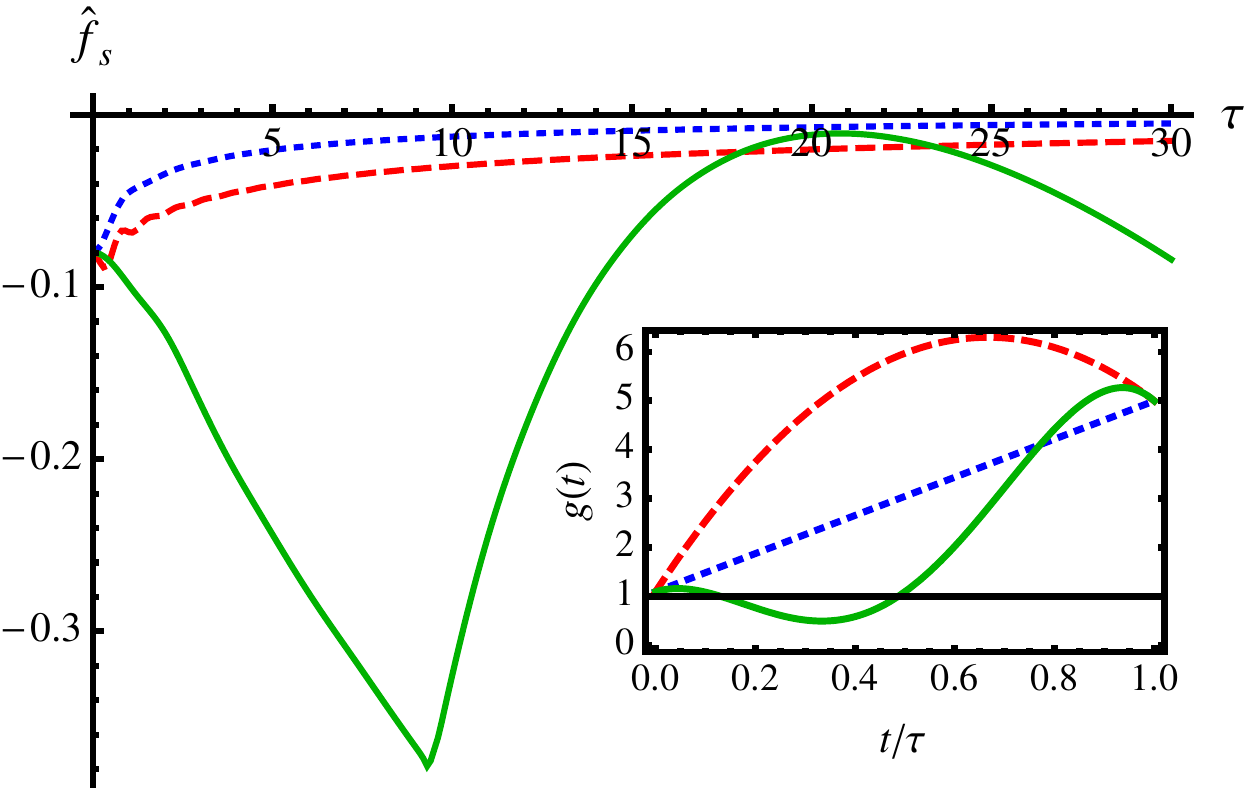}
\caption{(Color online) Plot of $\hat{f}_s$ [see Eq. (\ref{eq:log_fidelity})] for different protocols starting and ending in the same phase as a function of the duration $\tau$, with $g_0=1.1$ and $g_1=5$. The considered protocols are defined in Eq. (\ref{eq:protocols_ising}) and shown in the inset. In particular the dotted (blue) one is $g_{\rm lin}$, the dashed (red) one is $g_{\rm par}$ and the solid (green) one is $g_{\rm quart}$.}
\label{fig:fidelity_same_ising}
\end{center}
\end{figure}

\begin{figure}
\begin{center}
\includegraphics[width=\columnwidth]{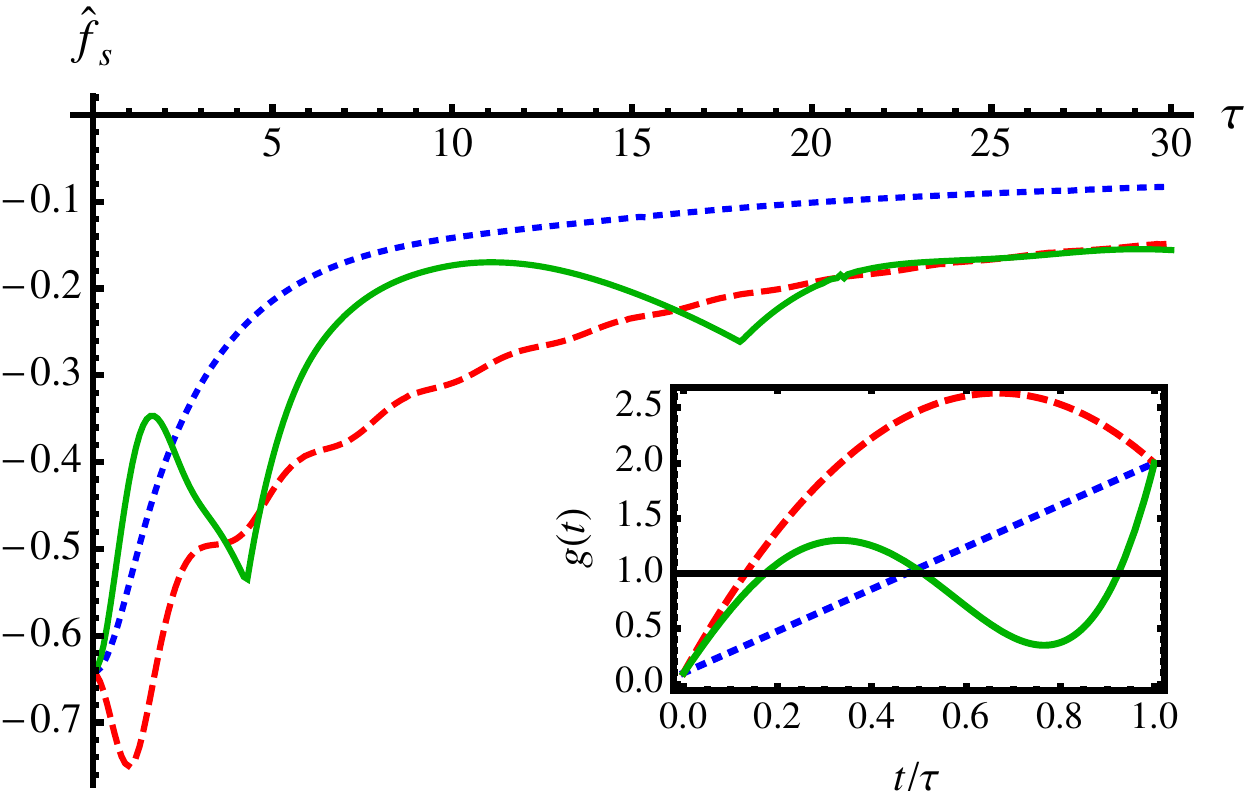}
\caption{(Color online) Plot of $\hat{f}_s$ (see Eq. (\ref{eq:log_fidelity})) for different protocols starting and ending in different phases as a function of the duration $\tau$, with $g_0=0.1$ and $g_1=2$. The considered protocols are defined in Eq. (\ref{eq:protocols_ising}) and shown in the inset. In particular the dotted (blue) one is $g_{\rm lin}$, the dashed (red) one is $g_{\rm par}$ and the solid (green) one is $g_{\rm quart}$.}
\label{fig:fidelity_ising_across}
\end{center}
\end{figure}

Using the result of the previous section, we can readily write down the cumulant distribution function for the full Ising chain
\beq
\frac{\ln G(s)}{L}=\int_0^\pi \frac{dk}{2 \pi} \ln\left(\frac{1+\abs{y_k(\tau)}^2 e^{-2 s \epsilon_k(\tau)} }{1+\abs{y_k(\tau)}^2}\right).
\label{eq:f_ising}
\eeq
Following Sec. \ref{sec:general}, we can identify the two contributions $f_c(s)=- \int \frac{dk}{2 \pi} \ln \left(1+\abs{y_k(\tau)}^2 e^{-2 s \epsilon_k(\tau)}\right)$ and $f_s=-1/2 f_c(0)$.

Again we notice that in the case of an adiabatic protocol the evolved state is the ground state of the final Hamiltonian, so $y_k(\tau)=0$, making $P(W)$ become a unique $\delta$-function peak at the origin, as expected. In the case of a sudden quench the state does not evolve, so $y_k(\tau)=y_k(0)$, which can be read off from Eq. (\ref{eq:mu_nu_k}) \cite{GS12}. 

If we consider a generic protocol, we can use Eq. (\ref{eq:a_b_ev}) to find an equation of  Riccati type that fully determines the function $y_k(\tau)$ and therefore the function $G(s)$,
\beq
i \frac{d}{dt} y_k(t)=-2 r_k(t) y_k(t)+s_k(t)\left(1-y_k(t)^2\right),
\label{eq:y_ev}
\eeq
with initial condition $y_k(0)=\frac{b_k(0)}{a_k(0)}$.

Contrary to the case of the free bosons, this equation has no diverging coefficients in the limit $k \rightarrow 0$ for protocols ending in the critical point, i.e., $g(\tau)=1$, but in the case of protocols across the critical point the initial function $y_k(0)$ diverges as $1/k^2$ for $k\rightarrow 0$. To avoid having divergent initial conditions one can define a new function $z_k(t)$ as
\beq
z_k(t)=\frac{1- \sign (g_0-g_1) y_k(t)}{1+\sign(g_0-g_1) y_k(t)},
\label{eq:z_def}
\eeq
so that the function $z_k(t)$ satisfies the  Riccati-like equation
\beq
i \frac{d}{dt} z_k(t)=r_k(t) (1-z_k^2(t))+2 s_k(t) \sign(g_1-g_0) z_k(t).
\label{eq:z_ev}
\eeq

As in the bosonic case, we start the comparison between different protocols by discussing the normalized log-fidelity $\hat{f}_s$. In Fig. \ref{fig:fidelity_same_ising} we show its behavior for different protocols as a function of the duration $\tau$, for $g_0=1.1$ and $g_1=5$, thus for protocols starting and ending in the paramagnetic phase. The protocols considered are linear, parabolic, and quartic, given by
\begin{subequations}
\beq
g_{\rm lin}=g_0+(g_1-g_0) t/\tau,
\eeq
\beq
g_{\rm par}=g_0+(g_1-g_0) (4 t/\tau-3 t^2/\tau^2),
\eeq
\beq
g_{\rm quart}= g_0 \sum_{n=1}^4 \rho_n (t/\tau)^n,
\eeq
\label{eq:protocols_ising}
\end{subequations}
with $\rho_n$ in the last protocol chosen in such a way that $g_{\rm quart}(t)$ is equal to $1/2$ for $t=1/3$ and $g_0$ for $t=1/2$. This protocol crosses the critical point and then returns in the paramagnetic phase. The actual values of the constants can be read in Appendix \ref{sec:coefficients} and the different plots are shown in the inset of Fig. \ref{fig:fidelity_same_ising} for $g_0=1.1$ and $g_1=5$.

From the figure we see that, as expected for both the linear and parabolic protocols the log-fidelity is essentially an increasing function of $\tau$ tending to zero (corresponding to fidelity going to one) for large $\tau$, with the parabolic protocol always giving a smaller value of the fidelity than the linear one. The quartic protocol has a very different behavior: it increases at the beginning and then decreases, displaying an oscillatory behavior (persistent for larger $\tau$) with an amplitude of the oscillations decreasing as $\tau$ increases. The qualitatively different behavior of this protocol has to be ascribed to the fact that it crosses the critical point and spends some time in the ferromagnetic phase before returning in the paramagnetic one.

In Fig. \ref{fig:fidelity_ising_across} $\hat{f}_s$ is plotted as a function of $\tau$ for the protocols defined in Eqs. (\ref{eq:protocols_ising}) with $g_0=0.1$ and $g_1=2$, i.e. protocols that start and end in different phases. The coefficients $\rho_n$ are now chosen in such a way that the protocol crosses the critical point three times. We see that for the linear protocol the fidelity is essentially an increasing function of $\tau$ with values that are larger than those in the same phase, and with an asymptotic value that appears to be different from zero. The parabolic protocol instead shows oscillations for small values of $\tau$ where it is possible to have a fidelity lower than what one gets for a sudden quench. Finally, the quartic protocol shows an oscillatory behavior and gives values of the fidelity always smaller than the sudden quench. 

We may now consider all the cumulants of the distribution $P(W)$ using Eq. (\ref{eq:cumulants}). The first cumulants are given by
\begin{subequations}
\beq
k_1= \langle W \rangle =2 L \int_0^\pi \frac{dk}{2 \pi} \frac{\epsilon_k(\tau) \abs{y_k(\tau)}^2}{1+\abs{y_k(\tau)}^2},
\eeq
\beq
k_2=\sigma^2=L \int_0^\pi \frac{d k}{2 \pi} \frac{4 \epsilon_k(\tau)^2\abs{y_k(\tau)}^2}{\left(1+\abs{y_k(\tau)}\right)^2},
\eeq
\beq
\frac{\sigma}{\langle W \rangle} \sim L^{-1/2},
\eeq
\beq
k_3= L \int_0^\pi \frac{d k}{2 \pi} \frac{8 \epsilon_k(\tau)^3 \left(\abs{y_k(\tau)}^4-\abs{y_k(\tau)}^2 \right)}{\left(1+\abs{y_k(\tau)}^2 \right)^3},
\eeq
\beq
\frac{k_3}{\sigma^3} \sim L^{-1/2}.
\eeq
\end{subequations}
The scaling with the size of the system $L$ of both $\ln G(s)$ and the cumulants is the same as in the case of free bosons. Therefore, also in this case it is convenient to define the intensive variable $w=W/L$, whose probability distribution has cumulants given by $\tilde{k}_n=L^{1-n} k_n$. Therefore, for large $L$ the distribution function $P(w)$ will be Gaussian with a mean equal to $k_1$ and variance given by $k_2/L$, which goes to zero for $L\rightarrow\infty$.

Figures \ref{fig:cumulants_same_ising} and \ref{fig:cumulants_across_ising} show the behavior of the first two cumulants per unit volume  for the protocols introduced previously that start and end in the same phase with $g_0=1.1$ and $g_1=5$ and for protocols that start and end in different phases with $g_0=0.1$ and $g_1=2$, respectively [the specific protocols considered are given by Eqs. (\ref{eq:protocols_ising})]. In the first case we see that the linear and parabolic protocols have cumulants that are essentially decreasing functions of $\tau$, with the parabolic cumulants always larger than the linear ones, while the quartic protocol shows cumulants with an oscillatory behavior. In the case of protocols starting and ending in different phases we see that the qualitative behavior for the linear protocol is almost the same as before, with the difference that the mean is not going to zero for large $\tau$.  This signals the fact that as one crosses the quantum critical point the adiabatic approximation breaks 
down. In the parabolic case we have oscillations for small $\tau$ and values always larger than the linear cumulants. Finally, in the case of the quartic protocol we see oscillations that are not as strong as in the case of protocols starting and ending in the same phase. 

\begin{figure*}
\begin{center}
\subfigure[\label{fig:mean_ising_same}]{\includegraphics[width=\columnwidth]{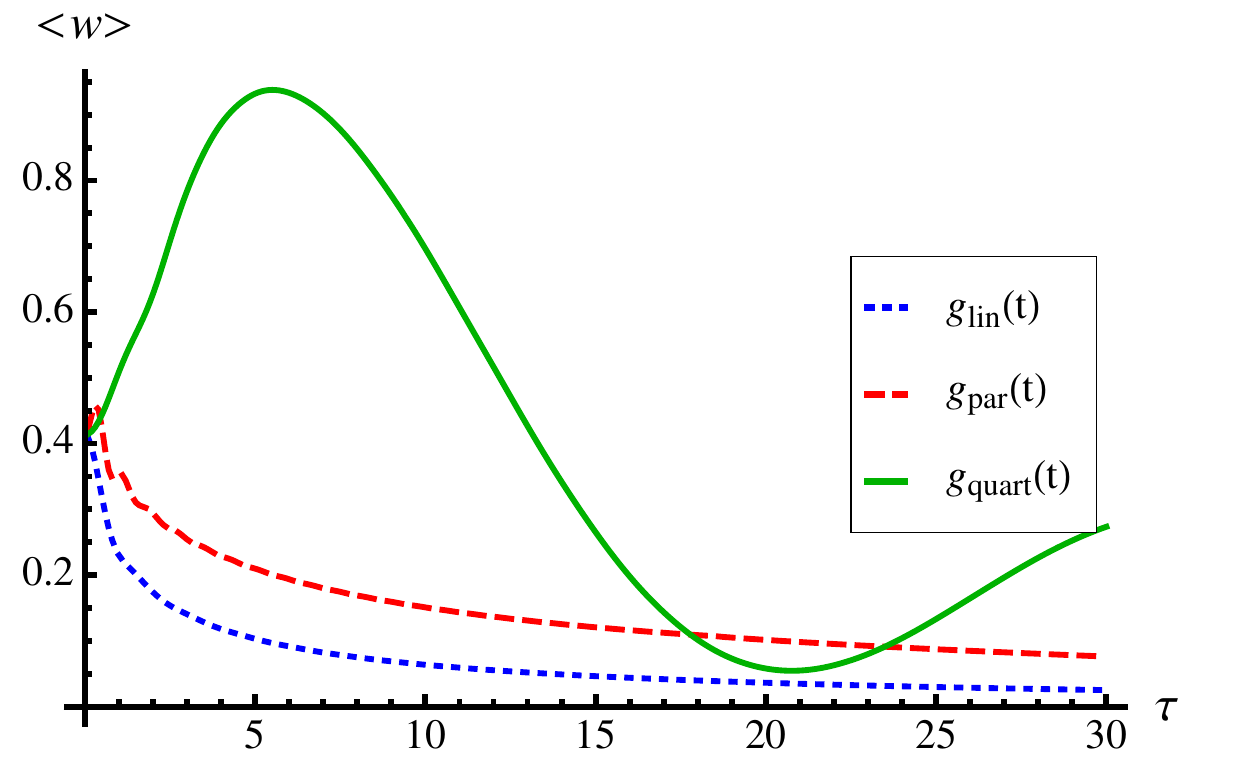}}
\subfigure[\label{fig:var_ising_same}]{\includegraphics[width=\columnwidth]{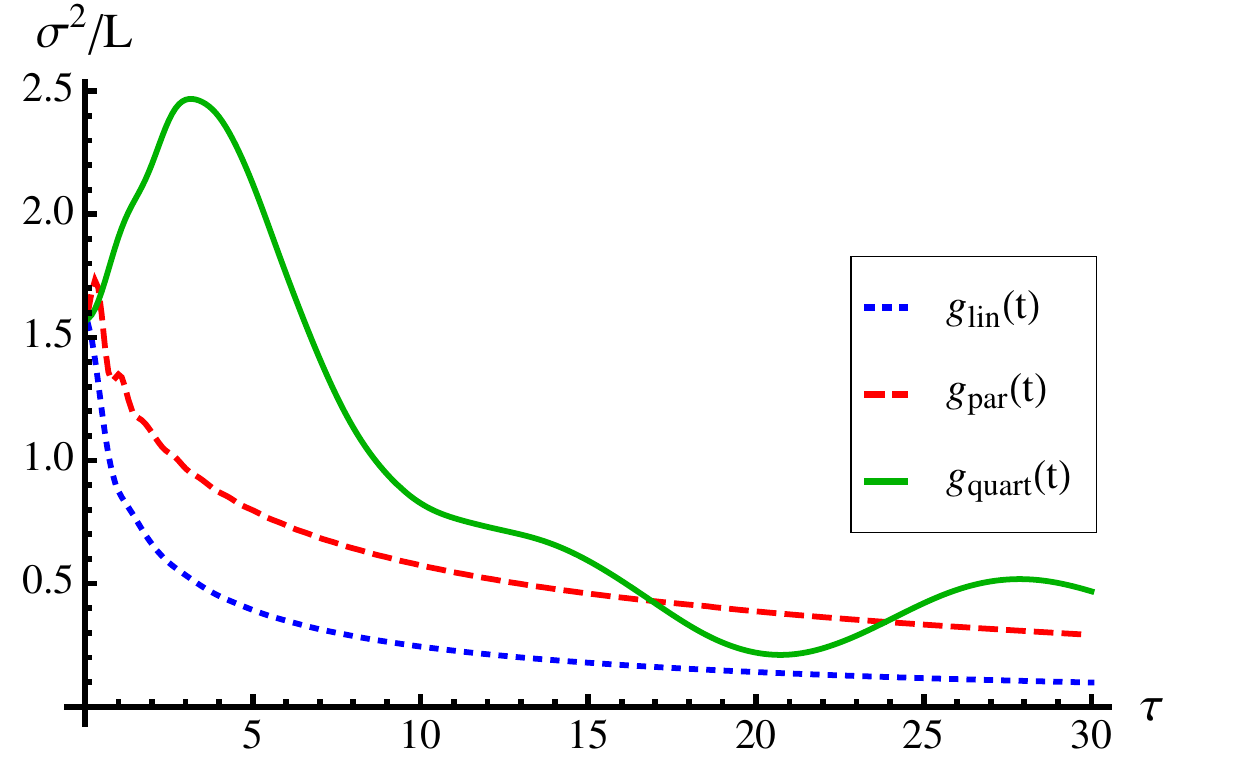}}
\caption{(Color online) Plot of (a) $\langle w \rangle$ and (b) $\sigma^2/L$   for the different protocols defined in Eqs. \ref{eq:protocols_ising} as a function of the duration $\tau$, with both $g_0=1.1$ and $g_1=5$ in the paramagnetic phase.}
\label{fig:cumulants_same_ising}
\end{center}
\end{figure*}

\begin{figure*}
\begin{center}
\subfigure[\label{fig:mean_ising_across}]{\includegraphics[width=\columnwidth]{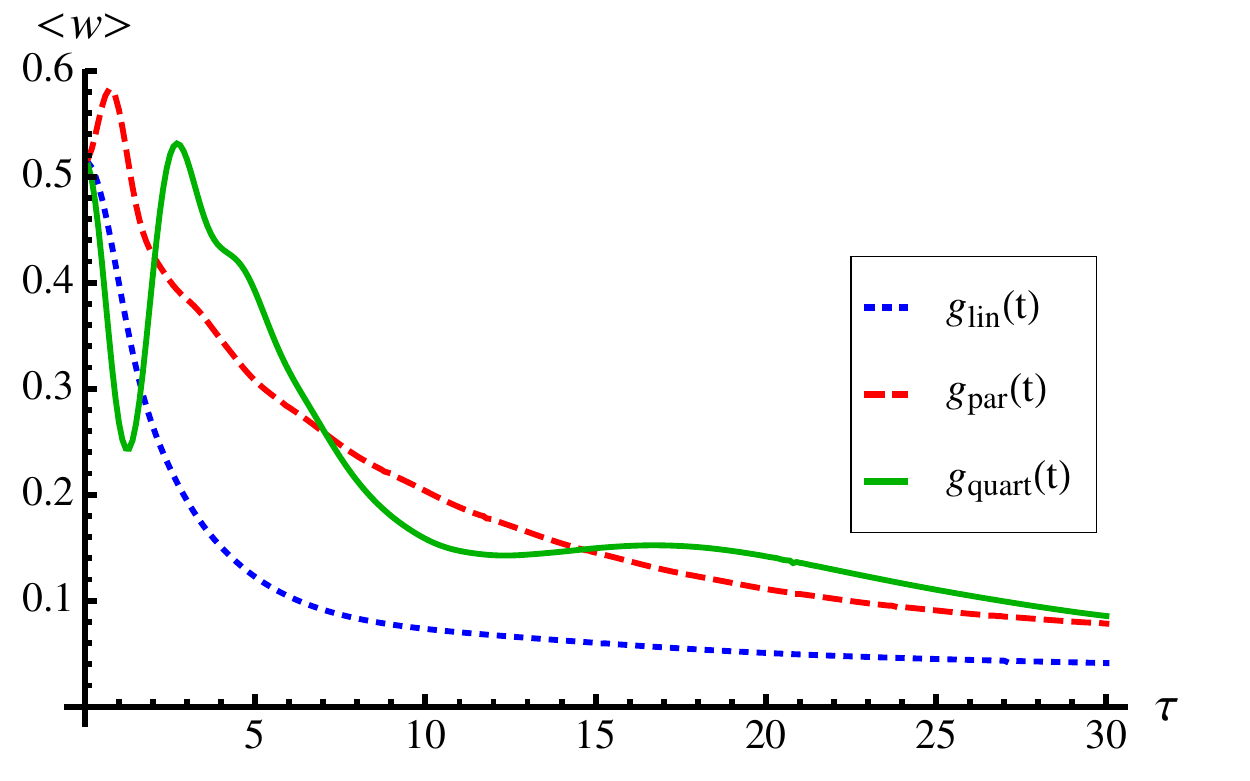}}
\subfigure[\label{fig:var_ising_across}]{\includegraphics[width=\columnwidth]{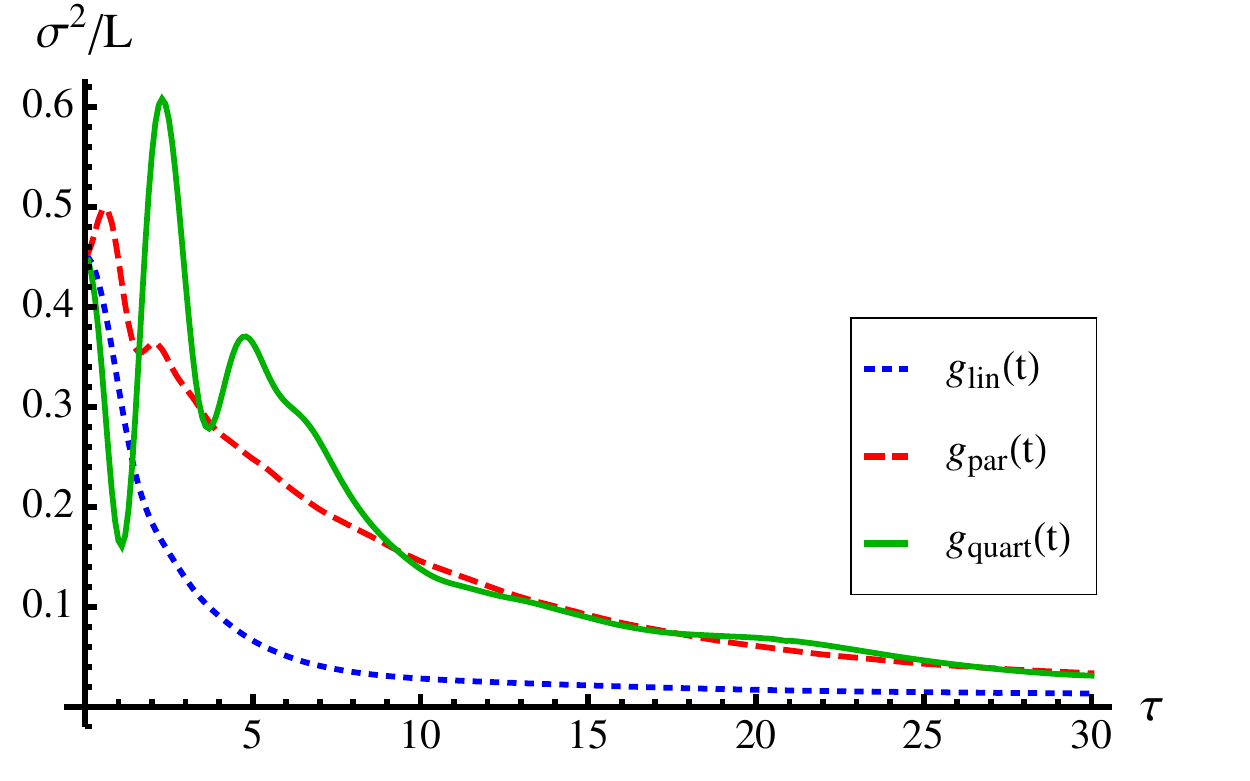}}
\caption{(Color online) Plot of (a) $\langle w \rangle$ and (b) $\sigma^2/L$  [see the definition below Eq. (\ref{eq:cumulants_boson})] for the different protocols defined in Eqs. \ref{eq:protocols} as a function of the duration $\tau$, with $g_0=0.1$ and $g_1=2$ in different phases.}
\label{fig:cumulants_across_ising}
\end{center}
\end{figure*}

To conclude this section we determine the behavior of the probability distribution of the work for small $W$ and study how its edge singularity is affected by the specifics of the chosen protocol. For this purpose we now consider the asymptotic behavior of $f_c(s)$, which in the case of a sudden quench has been analyzed in Ref. \onlinecite{Gambassi2011}. The asymptotic behavior of $f_c(s)$ for large $s$ is determined by the small-$k$ behavior of $\abs{y_k(\tau)}$, which depends crucially on whether the protocol crosses, does not cross, starts or ends exactly at the critical point. 

We start by considering the situation in which $g_0$ and $g_1$ are within the same phase, either paramagnetic or ferromagnetic. In this case, for an abrupt quench, $y_k(0)$ is an odd function of $k$ whose behavior for $k \rightarrow 0$ is 
\beq
y_k(0)= \frac{g_0-g_1}{2(g_0-1) (g_1-1)} k+ O(k^3).
\eeq
We now use Eq. (\ref{eq:y_ev}) to see if and how the small-$k$ behavior of $y_k$ is changed for a more general protocol. For this purpose we expand the function as a power series of $k$ as
\beq
y_k(t)=c_0(t)+c_1(t) k+c_2(t) k^2+O(k^3),
\label{eq:y_expansion}
\eeq 
with initial values $c_{2n}=0$ $\forall n$ and $c_1(0)=\frac{g_0-g_1}{2 (g_0-1) (g_1-1)}$. Then we use the evolution equation and the series expansion of Eq. (\ref{eq:r_s}) to obtain the equations for the coefficients $c_n(t)$, which for the first terms are given by
\begin{subequations}
\beq
i\frac{d}{dt} c_0(t)=2 [1-g(t)] \sign(g_1-1) c_0(t),
\eeq
\beq
\begin{split}
i \frac{d}{dt} c_1(t) = \,&\frac{g(t)-g_1}{\abs{1-g_1}} \left(1-c_0^2(t) \right)\\
&+2 \left(1-g(t) \right) \sign(g_1-1) c_1(t)
\end{split}
\eeq
\beq
\begin{split}
i\frac{d}{dt} c_2(t) =&2\, \frac{g_1-g(t)}{\abs{g_1-1}} c_0(t) c_1(t)+\frac{g(t)-g_1^2}{(g_1-1)^2} \sign(g_1-1) c_0(t) \\
& + 2 \left(1-g(t) \right) \sign(g_1-1) c_2(t)
\end{split}
\eeq
\label{eq:c_evolution_same}
\end{subequations}
We immediately see that $c_0(t)=0$ and $c_2(t)=0$ $\forall t$; it is possible to prove that the coefficients with even $n$ are all equal to zero. We may also write down explicitly the solution for $c_1(t)$ (which is the relevant one for the edge singularity). We obtain $\abs{y_k(\tau)}^2 \sim k^2 \abs{c_1(\tau)}^2$, with
\beq
\begin{split}
\abs{c_1(\tau)}^2=&\left(c_1(0)- \int_0^\tau\!\! ds \frac{g_1-g(s)}{g_1-1} \sin 2 \eta(s)\right)^2.\\
&+ \left(\int_0^\tau \!\!ds \frac{g_1-g(s)}{g_1-1} \cos 2 \eta(s)\right)^2,
\label{eq:coefficient_same}
\end{split}
\eeq
with $\eta(s)=\int_0^s[1-g(t)] dt$.

We therefore obtain that in general $\abs{y_k(\tau)}^2 \sim k^2$ for $k\rightarrow 0$  with a coefficient given by Eq. (\ref{eq:coefficient_same}). Therefore, the asymptotic behavior of $f_c(s)$ is
\beq
f_c(s) \simeq \frac{\abs{c_1(\tau)}^2}{8 \sqrt{\pi}} \left( \frac{\abs{1-g_1}}{s g_1} \right)^{3/2} e^{-2 s \abs{1-g_1}},
\eeq
implying for the distribution of the work,
\beq
\begin{split}
P(W) = &e^{-2 L f_S}\Bigl(\delta(W)+ L \frac{\abs{c_1(\tau)}^2}{4\pi}\\
&  \left(\frac{\abs{1-g_1}}{g_1}\right)^{3/2} \!\!\frac{\Theta(W-2 \abs{1-g_1})} {(W-2 \abs{1-g_1})^{-1/2}}+ \dots \Bigr).
\end{split}
\eeq
We notice that the specifics of the protocol appear only in the coefficient $c_1(\tau)$, while the exponent remains unaffected. We stress that the derivation above is valid also in the case in which $g(t)$ crosses the critical point at some instant of time, the only requirement being on the initial and final values. In the rather special case of a cyclic protocol, i.e., $g_1=g_0$, there are some minor modifications in the initial conditions, namely, $y_k(0)=0$, implying that the expansion coefficients of Eq. (\ref{eq:y_expansion}) at the initial time are $c_n(0)=0$ $\forall n$. As a result, in Eq. (\ref{eq:coefficient_same}) we have $c_1(0)=0$. 

\begin{figure}[t]
\begin{center}
\includegraphics[width=\columnwidth]{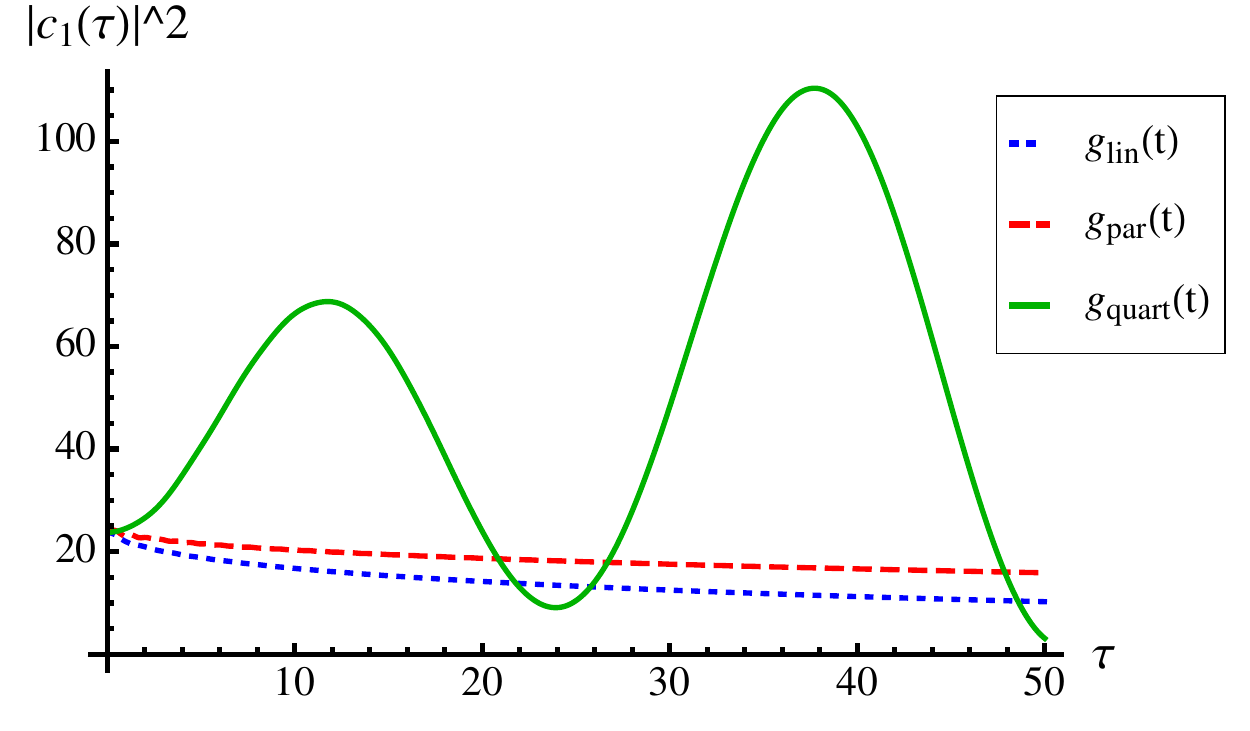}
\caption{Plot of the coefficient of the edge singularity $\abs{c_1(\tau)}^2$ for the different protocols starting and ending in the same phase defined in Eqs. (\ref{eq:protocols_ising}) with $g_0=1.1$ and $g_1=5$, as a function of $\tau$.}
\label{fig:edge_ising_same}
\end{center}
\end{figure}

In Fig. \ref{fig:edge_ising_same} we plot the value of $\abs{c_1(\tau)}^2$ as a function of $\tau$ for the protocols defined in Eqs. (\ref{eq:protocols_ising}) with $g_0=1.1$ and $g_1=5$. We see that for the parabolic and linear protocols this is a slowly decreasing function of $\tau$ with the values for the first protocol always larger than the others, while in the case of the quartic protocol we see an oscillatory behavior.

We now turn to the case of protocol starting at the critical point $g_0=1$ and ending in one of the two phases. In this case the equations for the coefficients of the series expansion (\ref{eq:y_expansion}) are still given by Eqs. (\ref{eq:c_evolution_same}), but with different initial conditions. Indeed we have
\beq
y_k(0)=\sign(1-g_1)+\frac{g_1+1}{2 (g_1-1)} k+O(k^2).
\label{eq:y_initial_fromcritical}
\eeq
The leading behavior for $k\rightarrow 0$ is now given by $c_0(\tau)$, whose value is
\beq
c_0(\tau)= \sign(1-g_1) e^{2i \sign(1-g_1) \int_0^\tau (1-g(s)) ds},
\eeq
from which we can immediately see that $\abs{c_0(\tau)}=1$ independently of the duration of the protocol and the final value of the transverse field $g_1$. So we conclude that in this case, not only the exponent, but also the coefficient of the edge singularity is the same for all the protocols. Indeed we have
\beq
f_c(s) \simeq \frac{1}{4 \sqrt{\pi}} \left( \frac{\abs{1-g_1}}{s g_1} \right)^{1/2} e^{-2 s \abs{1-g_1}},
\eeq
implying
\beq
\begin{split}
P(W)=&e^{-2 L f_s} \Bigl[  \delta(W)+\\
& \frac{L}{4 \pi} \sqrt{\frac{\abs{1-g_1}}{g_1}} \frac{\Theta(W-2\abs{1-g_1})}{\sqrt{W-2 \abs{1-g_1}}}+ \dots \Bigr].
\end{split}
\eeq
We now consider protocols ending at the critical point $g_1=1$. In this case the initial condition is given by 
\beq
y_k(0)=\sign(g_0-1)-\frac{g_0+1}{2 (g_0-1)} k +O(k^2),
\label{eq:y_initial_tocritical}
\eeq
which, apart from a minus sign, is the same as Eq. (\ref{eq:y_initial_fromcritical}) with the substitution $g_1 \rightarrow g_0$. However now also the small-$k$ behavior of $\epsilon_k(\tau)$ and so of $r_k(t)$ and $s_k(t)$ is changed in such a way that the equations for the coefficient of the expansion (\ref{eq:y_expansion}) are modified, becoming (up to order $k$)
\begin{subequations}
\beq
i \frac{d}{dt} c_0(t)=\left(g(t)-1 \right) \left(1-c_0^2(t) \right)
\eeq
\beq
i \frac{d}{dt} c_1(t)=2 \left(1-g(t)\right)c_1(t) c_0(t)-\left(g(t)+1\right)c_0(t),
\eeq
\end{subequations}
with initial conditions that can be read from Eq. (\ref{eq:y_initial_tocritical}). We immediately notice that $c_0= \pm 1$ is a stationary solution, so  also for quenches ending at the critical point both the exponent and the coefficient are independent of the choice of the protocol. In particular we have
\beq
f_c(s) \simeq \frac{1}{ 8 \pi s},
\eeq 
which implies
\beq
P(W) = e^{-2 f_s L} \Bigl[\delta(W)+L\frac{1}{8 \pi} \Theta(W)+ \dots \Bigr].
\eeq
Finally we consider the case of protocols starting in one phase and ending in the other one. In this case, as anticipated before, the behavior of $y_k(0)$ for small $k$ becomes singular, namely
\beq
y_k(0)= \frac{2 \abs{1-g_0} \abs{1-g_1}}{g_0-g_1} \frac{1}{k} + O(k),
\eeq
and  it is more convenient to consider the function $z_k$ defined in Eq.(\ref{eq:z_def}), whose behavior for $k\rightarrow 0$ at time $t=0$ is
\beq
z_k(0) =-1+\frac{\abs{g_0-g_1}}{\abs{g_0-1} \abs{g_1-1}} k+O(k^2).
\eeq
Also in this case we expand the function $z_k(t)$ in a power series 
\beq
z_k(t)=d_0(t)+d_1(t) k+d_2(t) k^2
\eeq
and use Eq. (\ref{eq:z_ev}) to determine the evolution of the coefficients, getting
\begin{subequations}
\beq
i\frac{d}{dt} d_0(t) = \left(g(t)-1\right) \sign(g_1-1)(1-d^2_0(t)),
\eeq
\beq
\begin{split}
i\frac{d}{dt} d_1(t)= &-2 \, \sign (g_1-1)\left(g(t)-1 \right) d_0(t) d_1(t)\\
&+2 \; \frac{g(t)-g_1}{\abs{1-g_1}} \sign(g_1-g_0)d_0(t)
\end{split}
\eeq
\end{subequations}
From this we derive that $d_0(t)=-1$ $\forall t$ and 
\beq
\begin{split}
d_1(\tau)= e^{2 i K(\tau)} \Bigl[&d_1(0) -2 i \int_0^\tau \sign(g_1-g_0) \frac{g_1-g(s)}{\abs{1-g_1}}\\
& e^{-2 i K(s)} \Bigr],
\end{split}
\eeq
where $K(t)=\sign(g_1-1) \eta(s)$, with $\eta(t)$ defined below Eq. (\ref{eq:coefficient_same}).

Returning to the function $y_k$, we have 
\beq
y_k(\tau)=-\frac{2}{d_1(\tau)}  \frac{1}{k}+O(1),
\eeq
so the leading behavior of $y_k(\tau)$ is still of the same type, implying that
\beq
\begin{split}
f_c(s) \simeq &\frac{1}{4 \pi} \left[\frac{4 \pi}{\abs{d_1(\tau)}} e^{-s \abs{1-g_1}} \right.\\
&\left. +\sqrt{\frac{g_1 s}{\abs{1-g_1}}} \Gamma(-1/2) \frac{4}{\abs{d_1(\tau)}^2} e^{-2 s \abs{1-g_1}} \right], 
\end{split}
\eeq
which for the probability distribution of the work gives, 
\beq
\begin{split}
& P(W)= e^{-2 L f_s} \Bigl[\delta(W)+\frac{L}{\abs{d_1(\tau)}} \delta \left(W-\abs{1-g_1}\right)\\
&+\frac{L^2}{\abs{d_1(\tau)}^2} \delta\left(W-2 \abs{1-g_1}\right)\\
&+\frac{L}{\pi} \frac{1}{\abs{d_1(\tau)}^2} \sqrt{\frac{g_1}{\abs{1-g_1}}} \frac{\Theta(W-2 \abs{1-g_1})}{(W-2 \abs{1-g_1})^{3/2}}\Bigr].
\end{split}
\eeq
Once again we notice that the choice of the protocol affects only the coefficient of the edge singularity, while the exponent is always the same.

In Fig. \ref{fig:edge_across_ising} we show the behavior of $2/\abs{d_1(\tau)}$ as a function of $\tau$ for the protocols defined in Eqs. (\ref{eq:protocols_ising}) with $g_0=0.1$ and $g_1=2$. We see that in the case of a linear protocol we have a decreasing function of $\tau$, while for the parabolic protocol we see oscillations for small $\tau$ that rapidly decrease in amplitude; for the quartic protocol we see an initial quite steep decrease and then small oscillations. The parabolic protocol always gives larger values at least for the time scale considered here.

\begin{figure}[t]
\begin{center}
\includegraphics[width=\columnwidth]{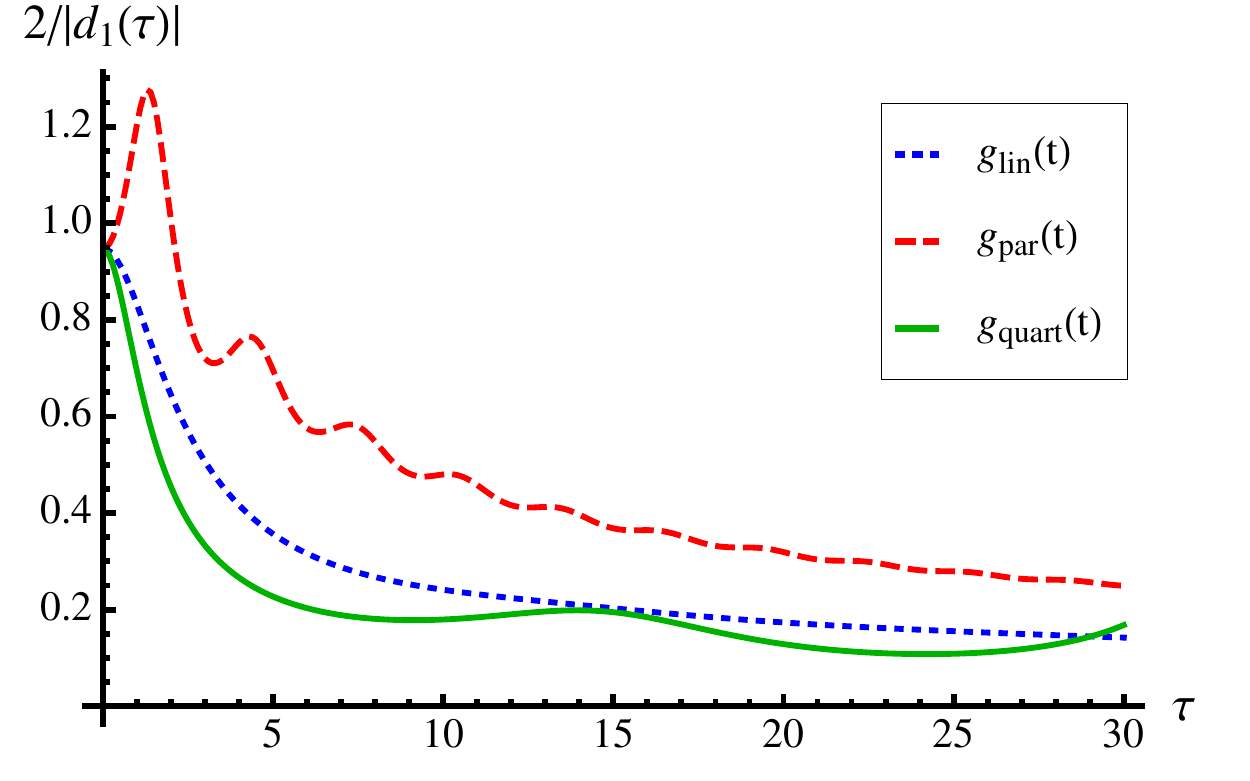}
\caption{Plot of the coefficient of the edge singularity $2/\abs{d_1(\tau)}$ for the different protocols starting and ending in different phases defined in Eqs. (\ref{eq:protocols_ising}) with $g_0=0.1$ and $g_1=2$, as a function of $\tau$.}
\label{fig:edge_across_ising}
\end{center}
\end{figure}

\section{Local protocol in the Ising chain}
\label{sec:local}

In this section we will again consider  the quantum Ising chain defined by the Hamiltonian (\ref{eq:ham_ising}), studying a new protocol in which  the transverse field $g$ starts from the critical value $g=1$ and is {\emph locally} changed in time. In order to analyze this problem we describe the system in the scaling limit by its corresponding conformal fiedl theory (CFT)~\cite{mussardo_09}, perturbed by a local mass term
\beq
H[m(t)]= -\frac{i}{2} \int dx \left[ \varphi \partial_x \varphi- \bar{\varphi} \partial_x \bar{\varphi} \right]+ i m(t) \bar{\varphi} \varphi_{\vert_{x=0}},
\label{eq:ham_local}
\eeq
where $\varphi$ and $\bar{\varphi}$ are two Majorana fermionic operators satisfying the commutation relations $\{\varphi(x),\varphi(x')\}=\{\bar{\varphi}(x), \bar{\varphi}(x')\}=\delta(x-x')$. We assume that the system is initially in its ground state. 

In this section we do not limit ourselves to the computation of the statistics of the work, but we will also consider the transverse magnetization and its correlations, produced by this kind of protocol. Part of the results presented in this section have been already been presented in Ref. \onlinecite{Smacchia2012}. 

The first step to solve this model and compute the statistics of the work is duplicating the theory using a trick first introduced by Itzynkson and Zuber \cite{zuber_itzykson}, i.e., introducing an additional pair of Majorana fermions $\chi$ and $\bar{\chi}$ described by the same Hamiltonian (\ref{eq:ham_local}) and anti-commuting with the original ones. From these two pairs of Majorana fermions we can then form two Dirac fermions
\beq
\psi_R = e^{-i \pi/4} \frac{ \varphi+i \chi}{\sqrt{2}}, \quad
\psi_L = e^{i \pi/4} \frac{ \bar{\varphi}+i \bar{\chi}}{\sqrt{2}},
\label{eq:RL_def}
\eeq
in terms of which the Hamiltonian reads
\beq
\begin{split}
H[m(t)] &= \int \!\!dx \left[ \psi^\dagger_R( -i \partial_x) \psi_R+ \psi^\dagger_L ( i \partial_x) \psi_L \right] \\
& + m \left( \psi^\dagger_L \psi_R + \psi^\dagger_R \psi_L\right)_{\vert_{x=0}}.
\end{split}
\eeq
In order to get a nonmixed mass term, we perform a nonlocal transformation \cite{kane_94}, defining
\beq
\begin{aligned}
\psi_+(x)&=\frac{(\psi_R(x)+\psi_L(-x))}{\sqrt{2}},\\
\psi_-(x)&=\frac{(\psi_R(x)-\psi_L(-x))}{\sqrt{2}i},
\label{eq:pm_def}
\end{aligned}
\eeq
so that we finally get
\beq
\begin{split}
H[m(t)]&=i\!\!\int \!\!\!dx \left[ \psi^\dagger_- \partial_x\psi_-  - \psi^\dagger_+ \partial_x \psi_+ \right]\\
&+m(t) \left[ \psi^\dagger_+ \psi_+ - \psi^\dagger_- \psi_- \right]_{\vert_{x=0}}.
\label{eq:ham_+-}
\end{split}
\eeq 
This transformed Hamiltonian describes two independent chiral modes that, since the Hamiltonian is quadratic, are completely characterized by the single-particle Hamiltonians $H_{+,-}= \mp i \partial_x \pm \delta(x) m(t)$. From this we can immediately write the equations of motion for $\psi_{+,-}$, which read
\beq
\left[i \partial_t\pm i \partial_x \right]\psi_{+,-}(x,t)= \pm \delta(x) m(t) \psi_{+,-}(x,t),
\eeq
whose initial condition is that $\psi_{\pm}(x,0)$ are free massless fermionic operators. These equations describe the scattering of a chiral field on a time-dependent $\delta$ potential:  for both $x>0$ and $x<0$ the field satisfies the free equation of motion, but after hitting the scatterer [in the region $x>0$ ($x<0$) for $\psi_+$ ($\psi_-$)], it gets a phase shift determined by the condition
\beq
\psi_{+,-}(0^{\pm},t)=\psi_{+,-}(0^{\mp},t) e^{\mp i m(t)}.
\eeq
From this we can derive the solution
\beq
\psi_{+,-}(x,t)= e^{\mp i m(t-\abs{x}) \theta(\pm x)} \psi_{+,-}(x\mp t,0),
\label{eq:psi_evolution}
\eeq
where $\psi_{+,-}(x,0)=\int \frac{dk}{\sqrt{2 \pi}} \, \hat{a}_{+,-}(k) e^{-\alpha \abs{k} /2} e^{\pm i k x}$, with $\alpha$ being the ultraviolet cutoff of the theory, and $\hat{a}_{+,-}$ the fermionic annihilation operators for the mode $k$.

We can now proceed to the computation of the average value of the transverse magnetization $\sigma^z \rightarrow  2 i \bar{\varphi}(x) \varphi(x)= i \left( \bar{\varphi}(x) \varphi(x)+ \bar{\chi}(x) \chi(x) \right)$ by first using Eqs. (\ref{eq:RL_def}) and (\ref{eq:pm_def}) to write it in terms of the Dirac operators $\psi_{+,-}$, getting
\begin{widetext}
\beq
\begin{split}
\mathcal{M}(x,t)&=\frac{1}{2} \left[\psi_+^\dagger(x,t) \psi_+(-x,t) + \psi_+^\dagger(-x,t) \psi_+(x,t)-\psi_-^\dagger(-x,t) \psi_-(x,t)-\psi_-^\dagger(x,t) \psi_-(-x,t)\right]\\
&-\frac{i}{2} \left[\psi^\dagger_+(x,t) \psi_-(x,t)- \psi^\dagger_+(-x,t) \psi_-(-x,t)+\psi^\dagger_-(-x,t) \psi_+(-x,t)-\psi^\dagger_-(x,t) \psi_+(x,t)\right].
\end{split}
\label{eq:mag_operator}
\eeq
\end{widetext}
We now compute the average of this operator over the initial state, i.e., the Dirac sea in which all the modes $k<0$ are occupied. We immediately see that the mixed terms in the second row average to zero, while the terms in the first row can be computed using Eq. (\ref{eq:psi_evolution}) and the mode expansion written below that equation, getting
\begin{subequations}
\beq
\langle \psi_+^\dagger(\pm x,t) \psi_+(\mp x,t) \rangle = \frac{e^{\mp i m(t-\abs{x})}}{2 \pi (\alpha \mp 2i x)}
\eeq
\beq
\langle \psi_-^\dagger(\pm x,t) \psi_-(\mp x,t) \rangle =\frac{e^{i m(t-\abs{x})}}{2 \pi (\alpha \pm 2 i x)}.
\eeq
\end{subequations}
Putting all the terms together we obtain the final result
\beq
\langle \mathcal{M} (x,t) \rangle = - \frac{ 2 \abs{x}}{\pi(4 x^2+\alpha^2)} \sin\left(m(t-\abs{x})\right).
\label{eq:transverse_magn}
\eeq 

\begin{figure}[tb]
\begin{center}
\subfigure[\label{fig:secondexamplea}]{\includegraphics[width=4.2 cm, height=3 cm]{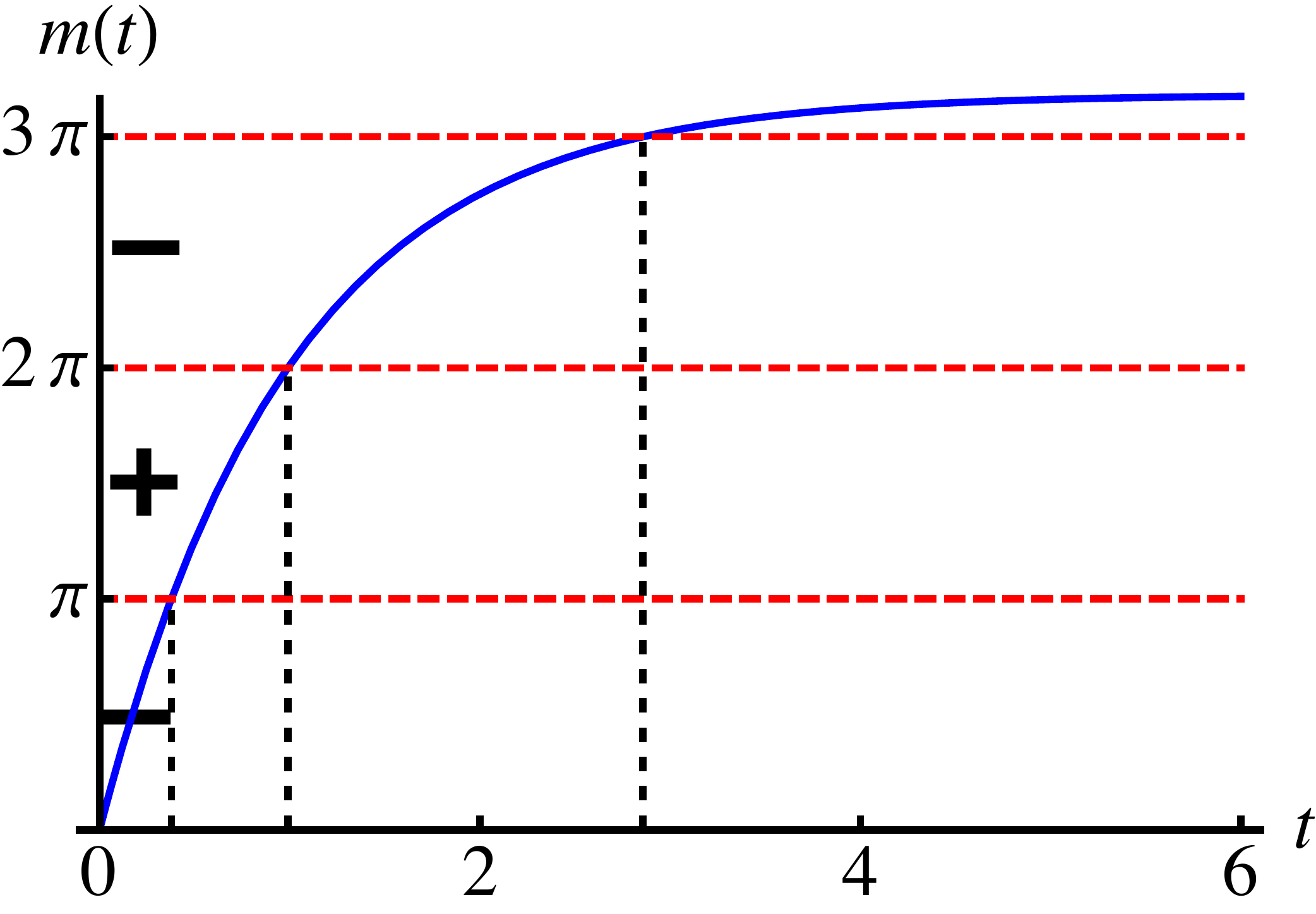}}
\subfigure[\label{fig:secondexampleb}]{\includegraphics[width=4.2 cm, height=3 cm]{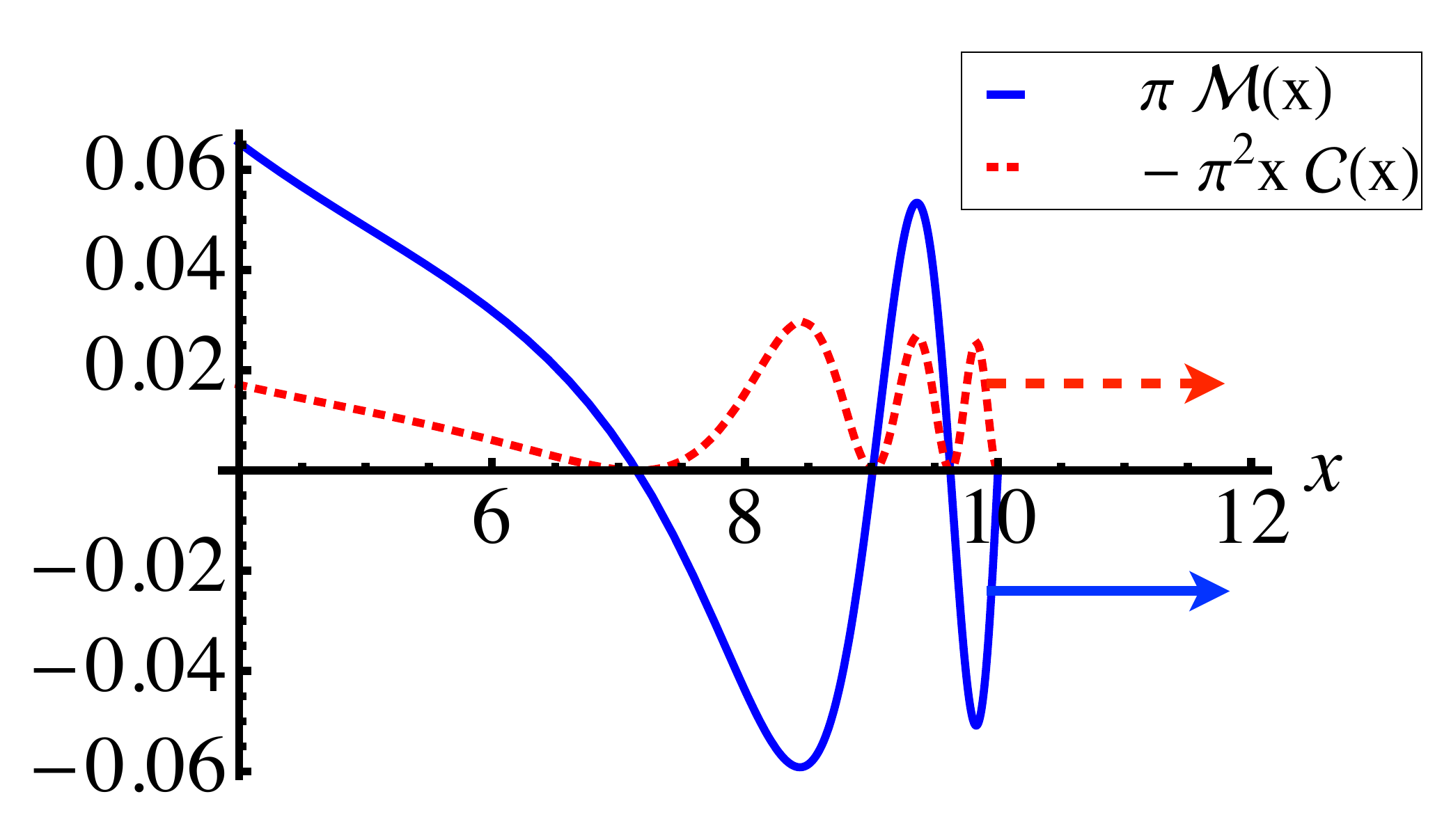}}
\subfigure[\label{fig:thirdexamplea}]{\includegraphics[width=4.2 cm, height=3 cm]{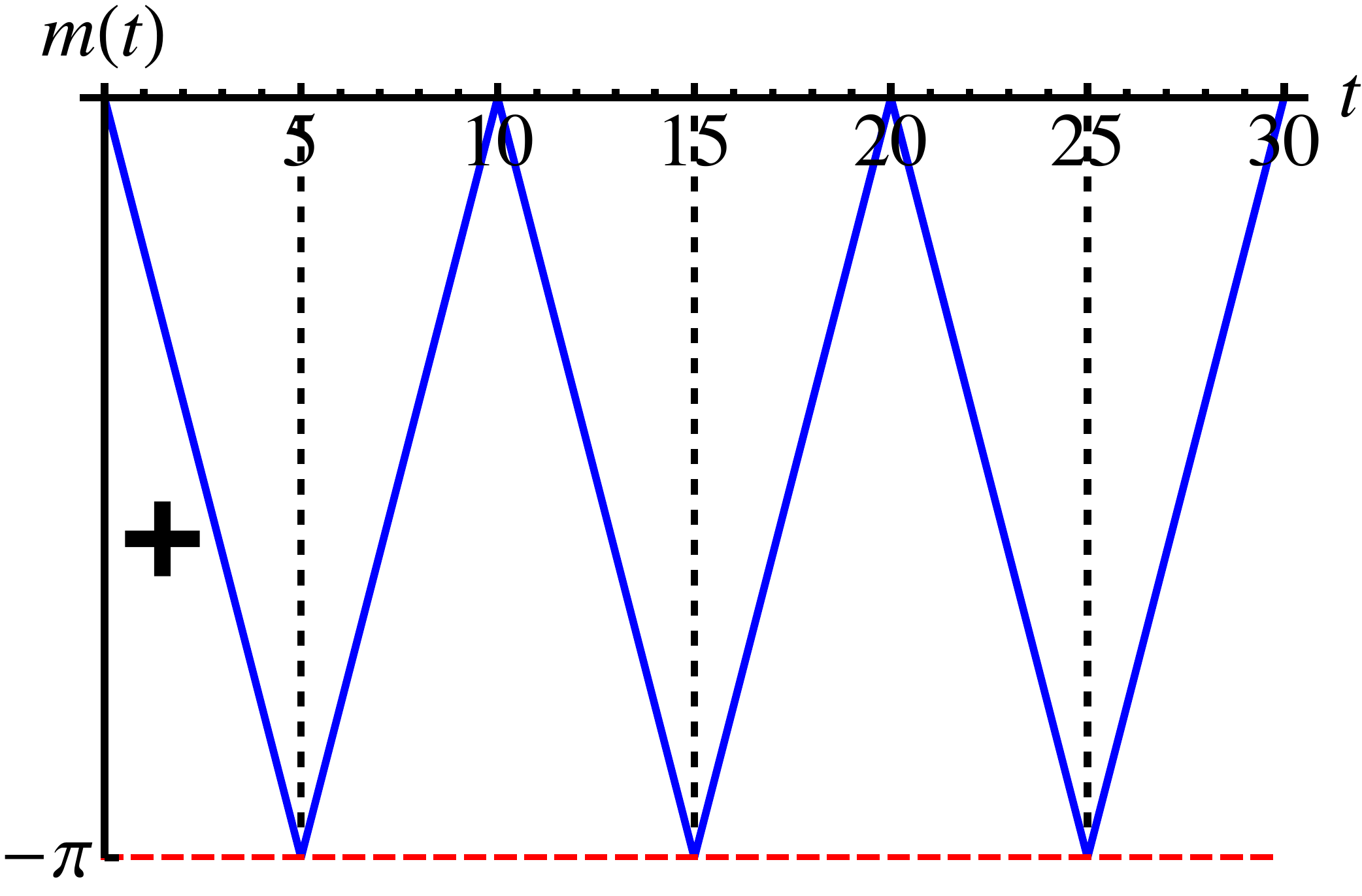}}
\subfigure[\label{fig:thirdexampleb}]{\includegraphics[width=4.2 cm, height=3 cm]{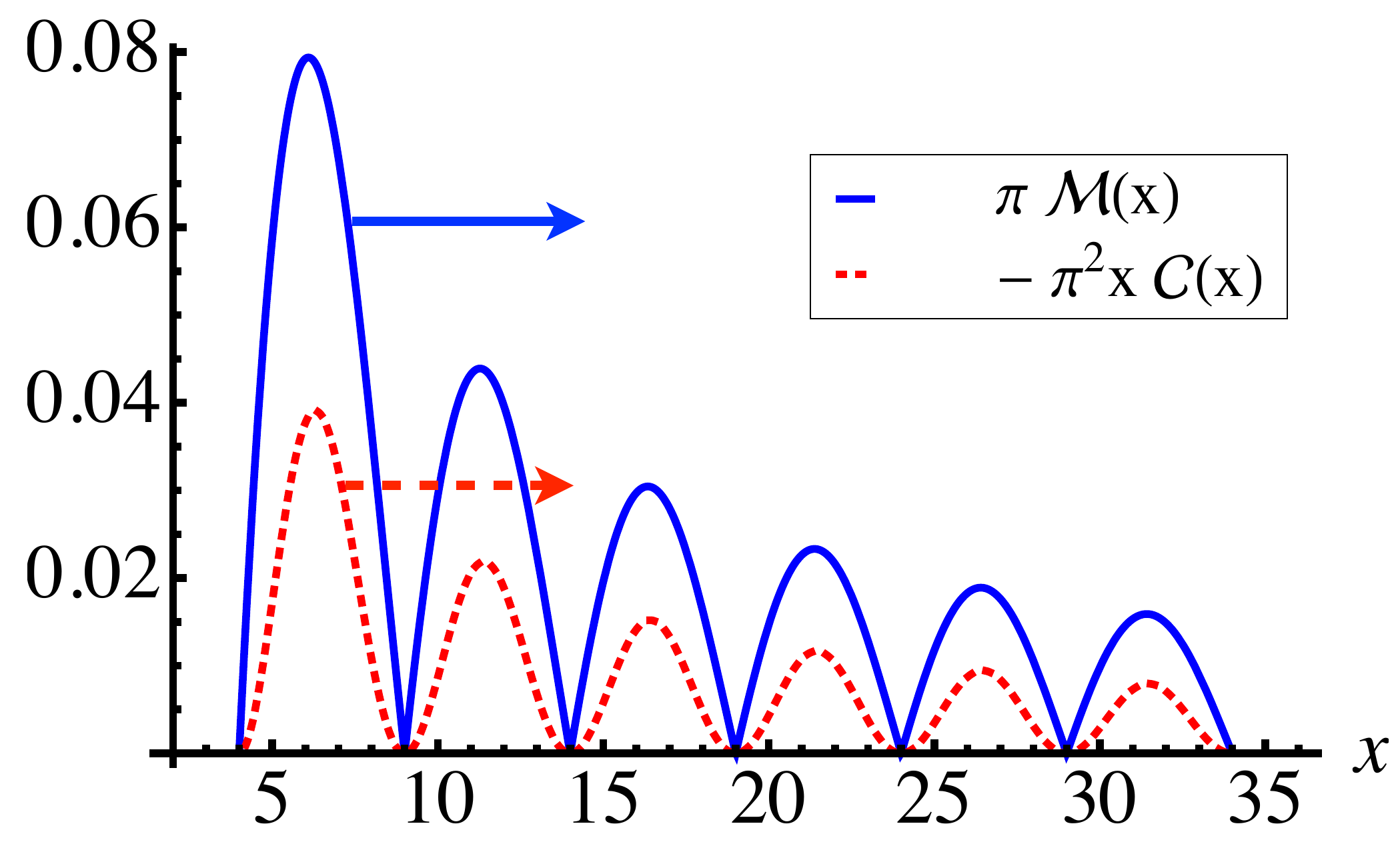}}
\caption{(Color online) Examples of magnetization and correlations profiles (right) for some specific protocol $m(t)$ (left). In (b) and (d) $t=10$.}
\end{center}
\end{figure}

This formula tells us that the local protocol $m(t)$ performed on the system causes the propagation at the velocity of light (which has been taken to be equal to $1$) of two identical magnetization profiles to the left and to the right of the origin. The amplitude of these profiles decreases with distance as $1/x$. 

Moreover, one can easily extract the qualitative features of the traveling signals. Indeed, the number of zeros is given by the number of times $m(t)$ crosses a value that is a multiple of $\pi$ and from the properties of the sine one can easily understand if the profile is positive or negative. As an example, in Fig. \ref{fig:secondexamplea} the protocol $m(t)= 10 (1-e^{-t}) \Theta(t)$ is analyzed. We conclude that the traveling profile will have three zeros, and a positive tail, since it asymptotically ends at a value between $3 \pi$ and $4 \pi$. Figure \ref{fig:secondexampleb} shows that these are indeed the features of the magnetization profile produced.  This simple understanding can be used to design protocols $m(t)$ producing a profile with the desired features. As an example, in Fig. \ref{fig:thirdexamplea} a protocol that produces six positive wave-packets with the same width is shown.

We can now use the same procedure to compute the connected correlations of the transverse magnetization at equal times $\langle \mathcal{M}(x,t) \mathcal{M}(x',t) \rangle_C$. Using Eq. (\ref{eq:mag_operator}) to compute the products of two magnetization operators at different points, one obtains $64$ terms, but the average over the initial state makes all the terms in which the number of $\psi_+$ and $\psi_-$ operators is different vanish, so one is left with $16$ relevant terms. At that point one can apply  Wick's theorem to decompose the products of four fermionic operators, then has to subtract the product of the average values at $x$ and $x'$ to get the connected correlations, and finally use Eq. (\ref{eq:psi_evolution}). More details can be found in Appendix \ref{sec:connected_correlation}. The final result of this procedure is
\begin{widetext}
\beq
\begin{split}
 &\langle \mathcal{M}(x,t) \mathcal{M} (x',t) \rangle_C= \cos\left(m(t-\abs{x})\right) \cos\left(m(t-\abs{x'})\right) \left[\frac{1}{2 \pi^2 \left[\alpha^2+(x-x')^2\right]} + \frac{\alpha^2}{2 \pi^2 (4x^2+\alpha^2) (4x'^2+\alpha^2)} \right]\\
&+  \sin\left(m(t-\abs{x})\right) \sin\left(m(t-\abs{x'})\right) \left[\frac{2 \Theta(x x') x x'}{\pi^2 \left[ \alpha^2+(x+x')^2 \right]\left[\alpha^2 +(x-x')^2 \right]}-\frac{2 \abs{x x'}}{\pi^2 (4 x^2+\alpha^2) (4x'^2+\alpha^2)} \right].
\end{split}
\label{eq:correlations}
\eeq
\end{widetext}
Since the magnetization profile is symmetric, the correlation between point $x$ and $-x$ is of particular interest. In particular the excess correlation $\mathcal{C}(x,t)=\langle \mathcal{M}(x,t) \mathcal{M}(-x,t) \rangle_C-\langle \mathcal{M}(x,0) \mathcal{M}(-x,0) \rangle_C$ is given by
\beq
\mathcal{C}(x,t)=\frac{1}{2 \pi^2 (4 x^2+\alpha^2)} \left( \cos \left(2 m(t-\abs{x}) \right)-1 \right).
\eeq
As in the case of the transverse magnetization, one may easily design the protocol $m(t)$ to give a certain desired correlation profile. In particular, the zeros of $\mathcal{C}(x,t)$ are the same as those of the magnetization, as one can also check in Fig.\ref{fig:secondexampleb} and \ref{fig:thirdexampleb} for specific protocols. More specifically, for every protocol $m(t)$ the excess correlations are always negative and travel through the system at the same speed of the magnetization, decreasing with the distance from the origin as $1/x^2$.

\subsection{Statistics of the work}
\label{sec:work_local}

We now turn to the computation of the statistics of the work done by this local protocol. In order to do so we will use a slightly different version of Eq. (\ref{eq:char_func}) and switch from the moment generating function to the characteristic function by performing the substitution $s \rightarrow -i \mu$. Thus we have
\beq
G(\mu)= {}_0 \langle 0 |e^{i \mu H^H[m(\tau)]} e^{-i \mu H[m(0)]} \ket{0}_0,
\eeq
where $H^H[m(\tau)]=U^\dagger(\tau) H[m(\tau)] U(\tau)$ is the final Hamiltonian in the Heisenberg picture, while $\ket{0}_0$ is always the initial ground state. Since we duplicated the theory, we will be actually computing $G^2(\mu)$.

The first step is to bosonize the Hamiltonian (\ref{eq:ham_+-}) using the usual formula,
\beq
\psi_{+,-}=\frac{1}{\sqrt{2 \pi \alpha}} e^{\pm i \sqrt{4 \pi} \phi_{\pm}(x)},
\label{eq:bos_form}
\eeq 
which gives (up to an irrelevant constant)
\beq
\begin{split}
H^H[\bar{m}]=&\int dx \left[\partial_x \phi_+(x,\tau)+\frac{\bar{m}}{2 \sqrt{\pi}} \delta(x) \right]^2\\
& +\left[\partial_x \phi_-(x,\tau)-\frac{\bar{m}}{2 \sqrt{\pi}} \delta(x) \right]^2.
\label{eq:bos_ham}
\end{split}
\eeq
where $\bar{m} \equiv m(\tau)$, and the bosonic operators $\phi_{+,-}$ are evolved with the full Hamiltonian until the final time $\tau$.

Then, using Eqs. (\ref{eq:psi_evolution}) and (\ref{eq:bos_form}) we can explicitly compute these evolved bosonic operators, which are given by
\beq
\phi_{+,-}(x,\tau)=\bar{\phi}_{+,-}(x,\tau)-\frac{m(\tau\mp x)}{\sqrt{4 \pi}} \Theta(\pm x),
\eeq
where $\bar{\phi}_{+,-}$ are instead bosonic field evolved with the free Hamiltonian. We can then use these expressions to write the Hamiltonian as
\beq
\begin{split}
H^H[\bar{m}]&=\int dx \left[ \partial_x \bar{\phi}_+(x,\tau)-\frac{\Theta(x)}{2 \sqrt{\pi}} \partial_x m(\tau-x) \right]^2\\
&+\left[ \partial_x \bar{\phi}_-(x,\tau)-\frac{\Theta(-x)}{2 \sqrt{\pi}} \partial_x m(\tau+x)\right]^2.
\end{split}
\eeq
We now look for an operator that shifts the derivatives of the fields $\bar{\phi}_{+,-}$ by the terms appearing in the preceding formula. To this aim the $\psi_+$ and $\psi_-$ operators can be treated independently. Let us then consider the $\psi_+$ operator and define $
\mathcal{U}_+(s)=e^{-i s \hat{A}}$, with
\beq
\hat{A}_+=\int_0^{+\infty} dy \, \bar{\phi}_+(y,\tau) \partial_y m(\tau-y).
\eeq
By using the commutation relation $[\bar{\phi}_+(x,t),\bar{\phi}_+(y,t)]=\frac{i}{4} \sign(x-y)$, we can derive the action of this operator on the derivative of the field $\bar{\phi}_+$, which is
\beq
\mathcal{U}^\dagger_+(s)\partial_x \bar{\phi}_+\mathcal{U}(s)=\partial_x \bar{\phi}_+(x,\tau)+\frac{1}{2} \, \Theta(x) \partial_x m(\tau-x) s,
\eeq
so  the choice $s=-1/\sqrt{\pi}$ gives the shift we were looking for.

We may now proceed in the same way for the operator $\psi_-$, so that at the end the unitary operator giving the shift we are looking for both the field $\bar{\phi}_+$ and the field $\bar{\phi}_-$ is $\mathcal{U}=\mathcal{U}_+ \mathcal{U}_-=e^{i / \sqrt{\pi}\hat{A}}$ with 
\beq
\hat{A}= \int_0^\infty dy \left( \phi_+(y-\tau,0)+\phi_-(\tau-y,0) \right) \partial_y m(\tau-y).
\eeq
Here we used the fact that $\bar{\phi}_{+,-}(y,t)=\phi_{+,-}(y \mp t)$, where from now on the field depending on just one variable is taken at the initial time $t=0$.

Putting everything together, we have that $H^H[\bar{m}]= \mathcal{U}^\dagger H[m(0)] \mathcal{U}$, which gives us
\beq
G^2 (\mu)= {}_0 \langle 0| \mathcal{U}^\dagger \mathcal{U} (\mu) \ket{0}_0,
\eeq
where in $\mathcal{U}(\mu)= e^{i \mu H[m(0)]} \mathcal{U} e^{-i \mu H[m(0)]}$ the bosons fields are evolved with the free Hamiltonian. Rewriting all in terms of the initial fields, we finally get
\begin{widetext}
\beq
\begin{split}
 G^2(\mu)&=\exp \left[\frac{1}{\pi} \int_0^{+\infty}dy \int_0^{+\infty} dy' \, \partial_y m(\tau-y) \partial_{y'} m(\tau-y') \left(\langle \phi_+(y-\tau) \phi_+(y'-\tau-\mu)\rangle \right. \right.\\
&+\langle\phi_-(\tau-y) \phi_-(\tau+\mu-y')\rangle -\frac{1}{2} \langle \phi_+(y-\tau)\phi_+(y'-\tau) \rangle-\frac{1}{2} \langle \phi_+(y-\tau-\mu) \phi_+(y'-\tau-\mu)\rangle\\
&\left. \left. -\frac{1}{2} \langle \phi_-(\tau-y) \phi_-(\tau-y')\rangle -\frac{1}{2} \langle \phi_-(\tau+\mu-y) \phi_-(\tau-y'+\mu)\rangle \right) \right].
\label{eq:g_bos2}
\end{split}
\eeq
\end{widetext} 
where $\langle \cdot \rangle= {}_0 \langle 0| \cdot | 0 \rangle_0$.

Now to compute the correlations of the bosonic fields we use their mode expansion
\beq
\phi_{\pm}(x)=\pm \!\int_0^{\pm \infty} \!\!\!\! dp \frac{e^{-\frac{\abs{\alpha}}{2} p}}{2 \pi \sqrt{2 \abs{p}}}  \left[e^{i px} \phi(p)+e^{-i px} \phi^\dagger(p) \right],
\eeq
with $\left[\phi(p),\phi^\dagger(p')\right]=2 \pi \delta(p-p')$, from which we obtain 
\beq
\begin{aligned}
\langle \phi_+(y-y'+\mu)\phi_+(0) \rangle &- \langle \phi_+(y-y') \phi_+(0)\rangle=\\
\frac{1}{4 \pi} \ln & \frac{\alpha-i (y-y')}{\alpha-i (y-y'+u)}.
\end{aligned}
\eeq
Finally, considering that $\langle \phi_-(x) \phi_-(y) \rangle= \langle \phi_+(y) \phi_+(x) \rangle$, we obtain the final result $G(\mu)= \exp[ F(\mu)]$, with
\beq
\begin{split}
F(\mu)=\frac{1}{4 \pi^2}& \int_{-\infty}^\tau \!\!\!\!dt  \int_{-\infty}^\tau \!\!\! dt'  \partial_t m(t) \partial_{t'} m(t')\\
& \ln\frac{\alpha-i (t-t')}{\alpha-i (t-t'+\mu)}.
\end{split}
\label{eq:g_local}
\eeq
From formula (\ref{eq:g_local}) we can compute all the cumulants of the distribution of the work using Eq. (\ref{eq:cumulants}). Doing so we get
\beq
\begin{split}
k_n= \frac{1}{4 \pi^2 n} \int_{-\infty}^{\tau} \!\!\!\!\! dt & \int_{-\infty}^{\tau} \!\!\!\!\! dt' \;  \partial_t m(t) \partial_{t'} m(t')\\
&\re \left[\frac{1}{\Bigl(\alpha-i (t-t')\Bigr)^n}\right].
\label{eq:cumulants_local}
\end{split}
\eeq
We immediately notice that, in contrast to what happens in the case of global protocols, and as anticipated before, the cumulants of $P(W)$ are not extensive, i.e. they are not proportional to the volume of the system. As a consequence, we do not have in general that the distribution tends to a Gaussian function in the limit of $L \rightarrow \infty$ with higher-order cumulants being suppressed by increasing power of the volume. 

We now show that the form of $P(W)$ for small $W$ is independent of the specifics of the protocol performed on the system. For this purpose, as already seen in the previous sections, we have to analyze the asymptotics of $G(\mu)$ for large $\mu$. When $m(\tau) \neq 0$, namely the final local mass is different from zero, we have
\beq
G(\mu) \simeq e^{\frac{B}{4 \pi}} (-i \mu)^{-\frac{\bar{m}}{4 \pi^2}},
\eeq
implying
\beq
P(W) \simeq B w^{\frac{\bar{m}^2}{4 \pi^2}-1},
\eeq
with $B=\int_{-\infty}^{\tau} \!\!\! dt \int_{-\infty}^{\tau} \!\!\! dt' \, \partial_t m(t) \partial_{t'} m(t') \ln [\alpha-i (t-t')]$.

Thus $P(W)$ displays an edge singularity with an exponent that depends only on the final value of the local mass but not on the way this value is reached. For small protocols ($\bar{m} < 2 \pi$) there is a power-law divergence, while for large protocols ($\bar{m} > 2 \pi$) $P(W)$ vanishes with a cusp. We observe that, as already anticipated in Sec. \ref{sec:general} and in contrast to what happens for global protocols, there is no $\delta$ peak at the origin, meaning that the probability that the final evolved state is in the ground state of the final Hamiltonian is zero. This is clearly a consequence of the Anderson orthogonality catastrophe \cite{anderson}.

\begin{figure}
\begin{center}
\subfigure[\label{fig:ex_w1}]{\includegraphics[width=4.2 cm, height=3.3 cm]{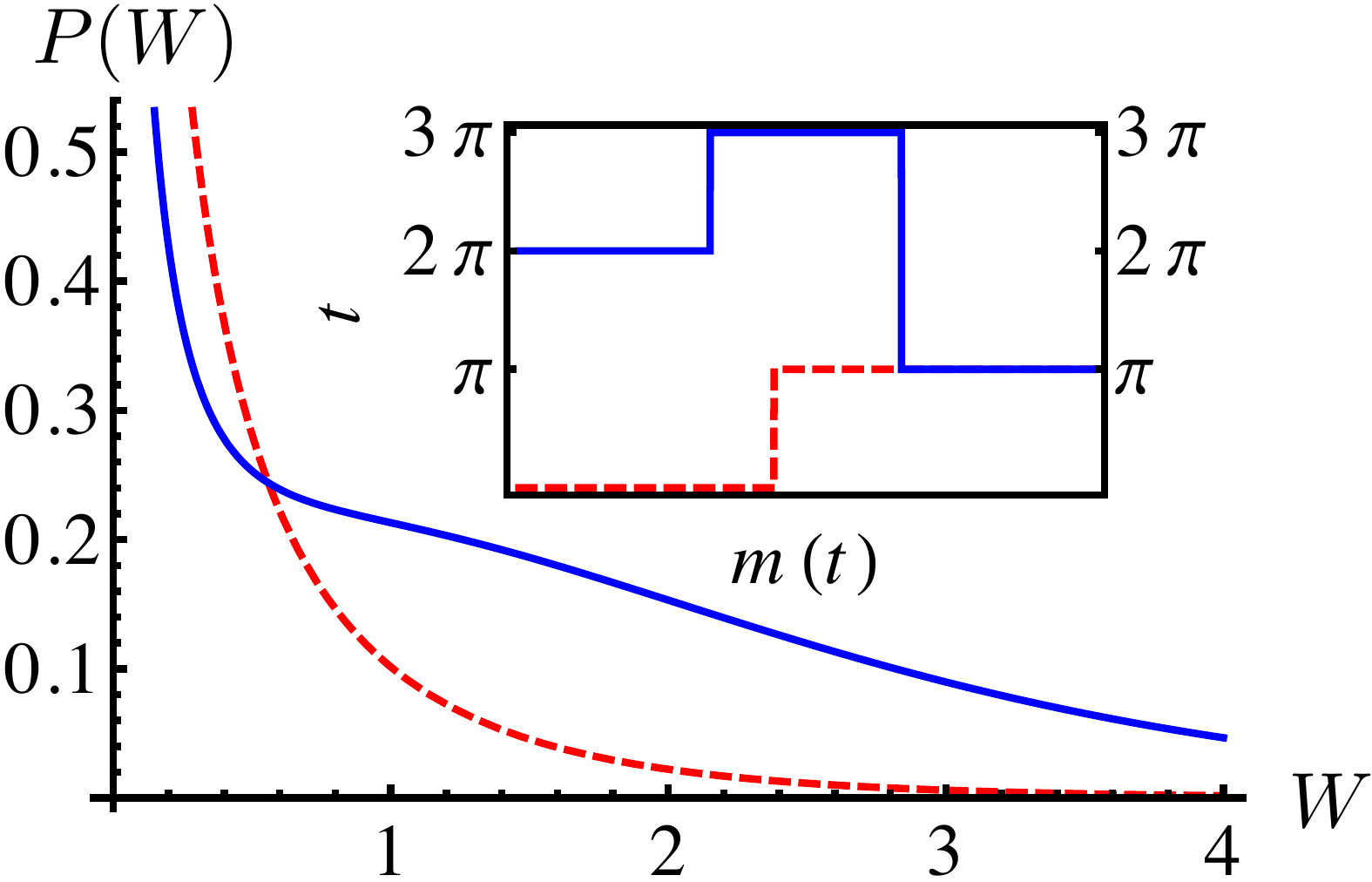}}
\subfigure[\label{fig:ex_w2}]{\includegraphics[width=4.3 cm, height=3.3 cm]{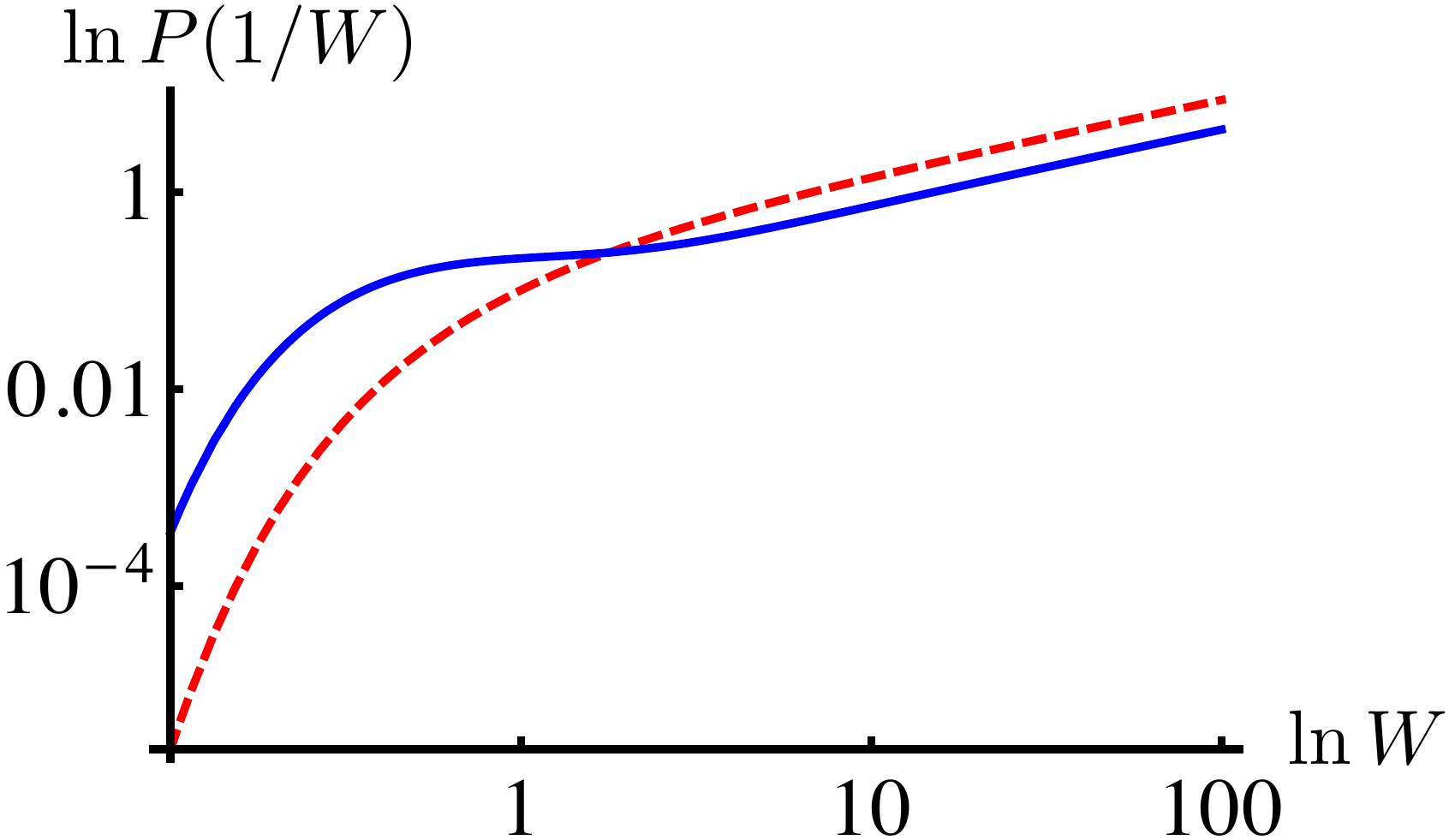}}
\caption{\footnotesize (Color online) (a) Probability distributions $P(W)$ for a nonmonotonic protocol solid (blue) line, i.e., a series of sudden quenches, and a sudden quench [dashed (red) line] ending at the same value of $m$ and shown in the inset. (b) Logarithmic plot of $P(1/W)$ for the same protocols as before. We set $\alpha=1$.}
\label{fig:example_work}
\end{center}
\end{figure}

We stress that this result, which may appear natural if one considers monotonic protocols (they all look like sudden quenches when the limit of large $\mu$ is taken), is  general: it holds independently of the shape of the protocol,  even in the case of nonmonotonic ones or in the case in which the critical point $m=0$ is crossed. We also note that, while in the case of global protocols (as seen in the previous sections) the spectral weight of the distributions tends to concentrate at a peak at high energies, so that observing the low-energy behavior becomes a rare event when the system size grows, for local protocols the low-energy part still retains a considerable spectral weight, making the power-law behavior likely to be observed. This is a consequence of the fact that, as already observed above, the cumulants of the distribution $P(W)$ are not extensive. The example of Fig. \ref{fig:example_work} clarifies the issue of both nonmonotonicity and observability. In Fig. \ref{fig:ex_w1} $P(W)$ is shown for 
a non monotonic protocol and a sudden quench to the same final value of the mass $m(\tau)$ (see the inset). One can see that in both cases the low energy part has a considerable spectral weight. From \ref{fig:ex_w2} one can see instead that the two protocols at low energy indeed behave as a power law with the same exponent.

The case of cyclic protocol, i.e., $m(\tau)=0$, is different. In this case the asymptotic behavior of the characteristic function becomes
\beq
G(u) \simeq e^{\frac{B}{4 \pi^2}}e^{C/\mu^2},
\eeq
with $C=\frac{1}{8\pi^2} \int_{-\infty}^{\tau} \!\!\! dt \int_{-\infty}^{\tau} \!\!\! dt' \, \partial_t m(t) \partial_{t'} m(t') (t-t'+i \alpha)^2 $, implying
\beq
P(W) \simeq e^{\frac{B}{4 \pi^2}}\left(\delta(W)+C \; W \right).
\eeq
In this case the $\delta$ peak is present, since orthogonality catastrophe no longer exists, and still the exponent of the edge singularity is independent of the details of the protocol, even from its final value, since now the regular part of $P(W)$ is always linear.

\begin{figure}[t]
\begin{center}
\includegraphics[width=\columnwidth]{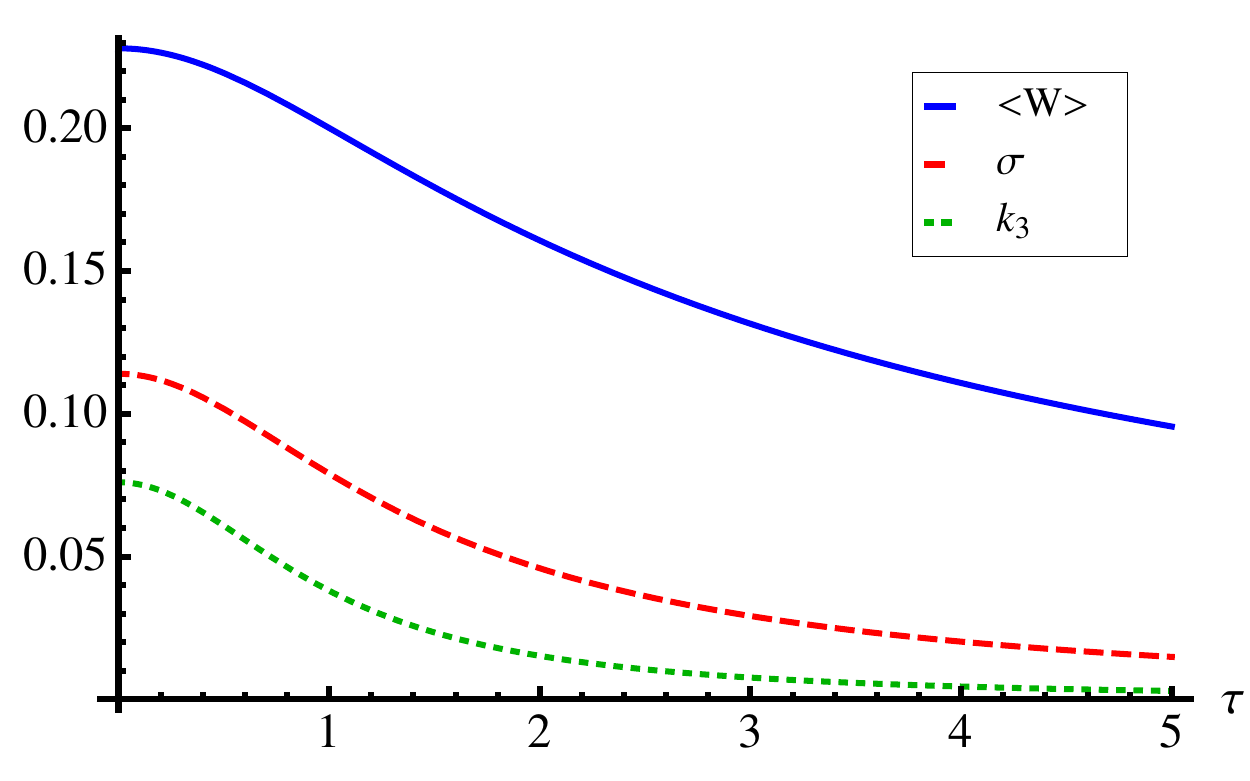}
\caption{\footnotesize (Color online) Cumulants for a linear ramp $m(t)$ reaching the final value $\bar{m}=3$ in a total time $\tau$. We set $\alpha=1$.}
\label{fig:cumulants_linear_local}
\end{center}
\end{figure}

In the case of non cyclic protocols we believe that our results can be extended to other one-dimensional systems and we propose a physical argument to support this claim. First, an asymptotic power-law behavior of $G(\mu)$ (and so the absence of a $\delta$ peak) has to be expected on the basis of the orthogonality catastrophe, which holds even if the final state is not the ground state of the initial Hamiltonian, since for a local protocol the former differs from the latter only for a finite number of excitations. Then, as explained in Sec. \ref{sec:general}, $G(\mu)$ can be interpreted as a partition function of a corresponding classical system on a strip of thickness $s$ after the Wick rotation $\mu \rightarrow i s$. The behavior for large $s$, which will determine the behavior of $P(W)$ for small $W$ is then expected to be determined by the $RG$ flow of the final state $\ket{\psi(\tau)}$ and the final Hamiltonian $H[m(\tau)]$.

The state should flow back to the initial critical state, since in its evolution only a finite number of excitations has been generated, while the flow of the Hamiltonian will depend on the nature of the defect, which can be marginal, irrelevant, or relevant. Therefore, the flow of the state should ensure the independence from the protocol, while the flow of the Hamiltonian should make the exponent universal in the usual sense of statistical mechanics. Moreover, in the case of a marginal defect (which is the one we explicitly considered here) this exponent should depend on the final strength of the defect (since the flow of the final Hamiltonian does), while in the case of a relevant perturbation we do expect this exponent to be completely independent of the protocol chosen and equal to $c/8-1$, where $c$ is the central charge, coming from the effect of a line of defect in a generic CFT \cite{cardy_2011}. An indication that this idea may be correct can be observed in the case of sudden quenches (or, 
equivalently, Fermi edge problem) in a Luttinger liquid \cite{gogolin_93,*affleck_94,*komnik_97,*furusaki_97}.  

We conclude this section studying some specific protocols in addition to the ones already considered in Fig. \ref{fig:example_work}. Let us start by considering a linear ramp reaching the final value $\bar{m}$. In this case, using formula (\ref{eq:g_local}) we have that
\beq
F_{\rm lin}(\mu)=\frac{\bar{m}^2}{\tau^2} \int^{\tau}_0 \!\!\!\!\! dx_1 \int^{\tau}_0 \!\!\!\!\! dx_2 \ln \frac{\alpha+i (x_1-x_2)}{\alpha+i (x_1-x_2+u)}.
\eeq
The integral can be done explicitly getting
\begin{widetext}
\beq
\begin{split}
 F_{\rm lin}(\mu)=\frac{\bar{m}^2}{4 \pi^2 t_f^2}&\left[\alpha^2 \ln \alpha-(\alpha-i\mu)^2 \ln (\alpha-i\mu) -\frac{(\alpha+i \tau)^2}{2}\ln(\alpha+i \tau)-\frac{(\alpha-i \tau)^2}{2} \ln (\alpha-i \tau)\right.\\
& \left. +\frac{(\alpha-i \mu+i \tau)^2}{2} \ln (\alpha-i \mu+i \tau)+\frac{(\alpha-i \mu-i \tau)^2}{2}\ln(\alpha-i\mu-i \tau) \right].
\label{eq:fi_linear}
\end{split}
\eeq
\end{widetext}
From this one can explicitly check that the asymptotic behavior is the one written above and can get all the cumulants of the distribution. For example, the first three cumulants are given by
\begin{subequations}
\beq
\langle W \rangle_{\rm lin}= \frac{\bar{m}}{4 \pi^2 \tau^2} \left[\alpha \ln \frac{\alpha^2}{\alpha^2+\tau^2}+2 \tau \arctan \frac{\tau}{\alpha}\right]
\eeq
\beq
\sigma^2_{\rm lin}=\frac{\bar{m}^2}{4 \pi^2 \tau^2} \ln \frac{\sqrt{\alpha^2+\tau^2}}{\alpha}
\eeq
\beq
k_{3,\rm lin}=\frac{\bar{m}^2}{4 \pi^2} \frac{1}{3 \alpha(\alpha^2+\tau^2)}
\eeq
\end{subequations}
In Fig. \ref{fig:cumulants_linear_local} we plot these cumulants as a function of $\tau$ for $\bar{m}=3$ and $\alpha=1$.

We finally consider an example of a cyclic protocol, namely, a parabolic protocol of total duration $\tau$ reaching its maximal amplitude of $k (\tau/2)^2$ at $t=\tau/2$. Using the general formula (\ref{eq:g_local}) we get
\beq
F_{\rm par}(\mu)=\frac{k^2}{\pi^2} \int_{-\tau/2}^{\tau/2} dt \int_{-\tau/2}^{\tau/2} dt' t t' \ln \frac{\alpha-i (t-t')}{\alpha-i (t-t'+\mu)}, 
\eeq
which can be computed to obtain,
\begin{widetext}
\beq
\begin{split}
F(\mu) &=\frac{k^2}{\pi^2} \biggl[ \frac{\alpha}{12}(\alpha^3+3 \tau^2 \alpha+i \tau^3) \arctanh 
\left(\frac{\tau}{2 i \alpha-T} \right)-\frac{\alpha}{12} (\alpha^3+3 \tau^2 \alpha-i \tau^3) \arctanh \left( \frac{\tau}{2i \alpha+\tau} \right)\\
&+ \frac{1}{12} \tau^3 \alpha \arctan \left(\frac{\tau}{\alpha} \right)+\frac{\alpha^2 \tau^2}{24}-\frac{1}{12} (\alpha-i\mu) \left[(\alpha-i\mu)^3+3 \tau^2 (\alpha-i\mu) +i \tau^3 \right]\\
&\arctanh \left( \frac{\tau}{2 \mu- \tau+2 i \alpha} \right) +\frac{1}{12} (\alpha-i\mu) \left[(\alpha-i\mu)^3+3 \tau^2 (\alpha-i\mu) -i \tau^3 \right] \arctanh \left( \frac{\tau}{2i \alpha+\tau+2 u} \right)\\
&-\frac{1}{12} \tau^3 (\alpha-i\mu) \arctan \left( \frac{\tau}{ \alpha-i\mu} \right)- \frac{(\alpha-i\mu)^2}{24} \tau^2 \biggr].
\label{eq:f_par}
\end{split}
\eeq
\end{widetext}
Also in this case one can check that indeed the asymptotic behavior is the same we obtained above and one can compute all the cumulants of the distribution $P(W)$. In particular the first two are given by
\begin{subequations}
\beq
\begin{split}
\langle W \rangle_{\rm par} = &\frac{k^2}{6 \pi^2} \left[\tau^2 \alpha+\tau^3 \arctan\left(\frac{\tau}{\alpha} \right) \right. \\
& \left. -(3 \tau^2 \alpha+2 \alpha^3) \ln \frac{\sqrt{\alpha^2+\tau^2}}{\alpha} \right],
\end{split}
\eeq
\beq
\sigma^2_{\rm par}=\frac{k^2}{12 \pi^2} \left[(\tau^2+2 \alpha^2) \ln \frac{\sqrt{\alpha^2+\tau^2}}{\alpha}-\tau^2 \right].
\eeq
\end{subequations}
In Fig. \ref{fig:cumulants_parabolic_local} we show the behavior of these two cumulants as a function of the total time $\tau$ for $k=1$ and $\alpha=1$.

\begin{figure}
\begin{center}
\includegraphics[width=\columnwidth]{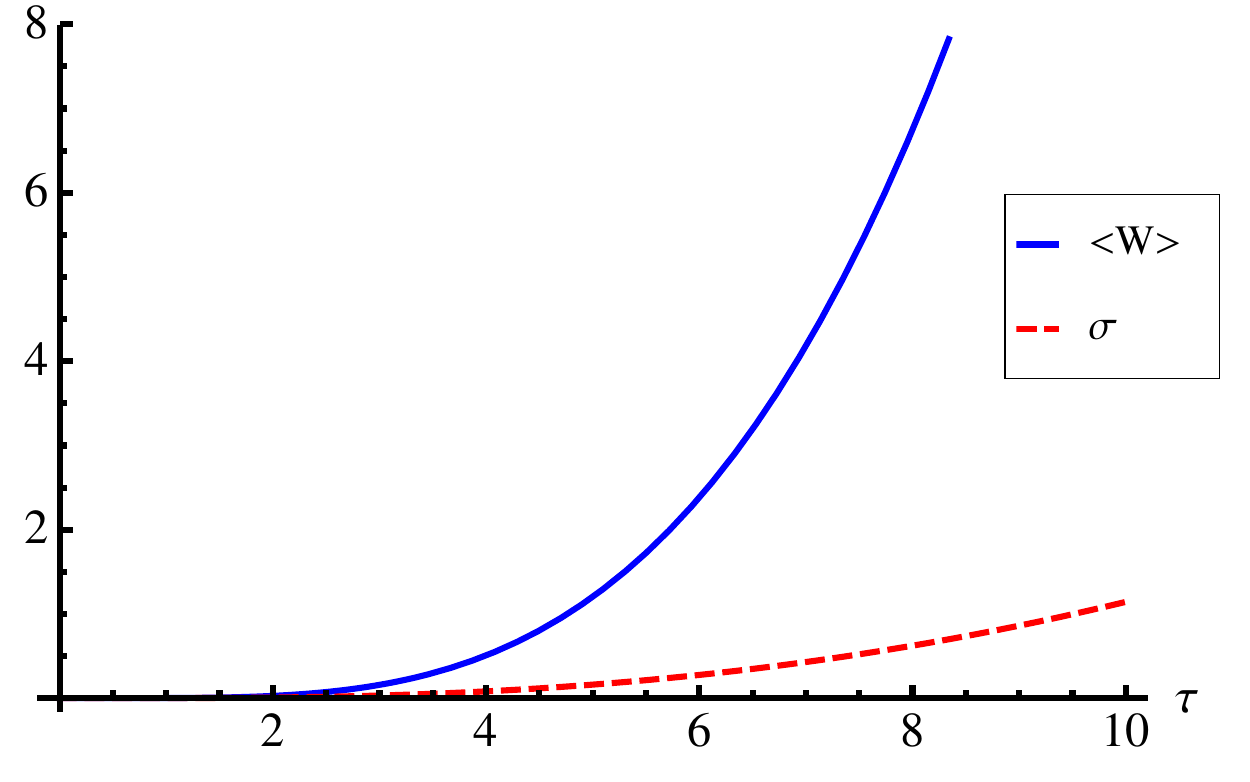}
\caption{\footnotesize (Color online) Cumulants for a parabolic protocol $m(t)$, which returns to $m=0$ in a total time $\tau$ and reaches its maximum value of $\tau^2/4$. We set $\alpha=1$.}
\label{fig:cumulants_parabolic_local}
\end{center}
\end{figure}

\section{Conclusions}
\label{sec:conclusions}

In this work we studied the statistics of the work done by performing a generic time-dependent protocol on some integrable Hamiltonians, which can be represented as free bosons or fermions, for both global and local quenches.  We provided exact expressions for this quantity in the case of a free bosonic field theory with relativistic dispersion relation, in which the mass is globally changed in time, and in the case of a global or local change of the transverse field in the one-dimensional Ising chain. These results allow us to compute, among other things, the fidelity of the final state and all the cumulants of the distribution, which can be used in optimization procedures with or instead of the fidelity.

Moreover, we found that the exponent of the edge singularity that is present in the low-energy part of $P(W)$ turns out to be independent of the specifics of the protocols and depends only on its general properties. For the global protocols of Sec. \ref{sec:bosons} and \ref{sec:ising_global} it depends only on the position of the starting and ending points with respect to the critical point of the system, namely, if they are both in the same phase, in different phases, or one of the two is the critical point itself. We found that for protocols starting from or ending at the critical point also the coefficient of the edge singularity is universal when it is rescaled with respect to an overall factor determined by the fidelity, both for the bosons and for the Ising model. In the case of the local protocol of Sec. \ref{sec:local} the exponent is determined only by the final value of the  local mass. We also provided a physical argument that suggests a possible generalization of our result to other systems.

Even though at the moment there is no experimentally viable technique for measuring the distribution of the work $P(W)$ for many-body quantum systems, suitable extensions of recent proposals might provide access to its generating function \cite{Dorner2013,*Mazzola2013}.  

\appendix

\section{Computation of the connected correlations of the transverse magnetization}
\label{sec:connected_correlation}

In this appendix we give additional details about the computation of the connected correlations of the transverse magnetization. As already mentioned,  when we multiply two magnetization operators (\ref{eq:mag_operator}) at points $x$ and $x'$ we get $64$ terms; however we can disregard terms in which the number of creation operators is not equal to the number of annihilation operators for at least one of the two species $\psi_+$ and $\psi_-$ of chiral fermions. Doing so, we are left with 16 relevant terms
\begin{widetext}
\beq
\begin{split}
 & \left\langle \mathcal{M}(x,t) \mathcal{M}(x',t) \right\rangle= \frac{1}{4}\left\langle \psi^\dagger_+(x)\psi_+(-x)\psi^\dagger_+(x')\psi_+(-x')+\psi^\dagger_+(x) \psi_+(-x)\psi^\dagger_+(-x')\psi_+(x') \right. \\
&+ \psi^\dagger_+(-x)\psi_+(x)\psi^\dagger_+(x')\psi_+(-x')+\psi^\dagger_+(-x) \psi_+(x)\psi^\dagger_+(-x')\psi_+(x')+\psi^\dagger_-(x)\psi_-(-x)\psi^\dagger_-(x')\psi_-(-x')\\
& +\psi^\dagger_-(x) \psi_-(-x)\psi^\dagger_-(-x')\psi_-(x')+\psi^\dagger_-(-x)\psi_-(x)\psi^\dagger_-(x')\psi_-(-x')+\psi^\dagger_-(-x)\psi_-(x)\psi_-^\dagger(-x')\psi_-(x')\\
&-\psi^\dagger_+(x)\psi_-(x)\psi^\dagger_-(-x')\psi_+(-x')+\psi^\dagger_+(x)\psi_-(x)\psi^\dagger_-(x')\psi_+(x')+\psi^\dagger_+(-x)\psi_-(-x)\psi^\dagger_-(-x')\psi_+(-x')\\
&- \psi^\dagger_+(-x)\psi_-(-x)\psi^\dagger_-(x')\psi_+(x')-\psi^\dagger_-(-x)\psi_+(-x)\psi^\dagger_+(x')\psi_-(x')+\psi^\dagger_-(-x)\psi_+(-x)\psi^\dagger_+(-x')\psi_-(-x')\\
&\left.+\psi^\dagger_-(x)\psi_+(x)\psi^\dagger_+(x')\psi_-(x')-\psi^\dagger_-(x)\psi_+(x)\psi^\dagger_+(-x')\psi_-(-x') \right\rangle.
\end{split}
\eeq
\end{widetext}
Here and in the following we will not explicitly write the time dependence of the fermionic operators.

The next step is to decompose the averages of products of four fermionic operators using the Wick theorem and then subtracting the product of the averages of the magnetization at the points $x$ and $x'$ in order to get the connected correlation. If we do so we get
\begin{widetext}
\beq
\begin{split}
&\langle \mathcal{M} (x,t)\mathcal{M} (x',t) \rangle_C= \frac{1}{4} \left[ \langle \psi^\dagger_+(x) \psi_+(-x') \rangle \langle \psi_+(-x) \psi^\dagger_+(x') \rangle +\langle \psi^\dagger_+(x) \psi_+(x') \rangle \langle \psi_+(-x) \psi^\dagger_+(-x') \rangle +\langle \psi^\dagger_+(-x) \psi_+(-x') \rangle  \right.\\
&\langle \psi_+(x) \psi^\dagger_+(x') \rangle + \langle \psi^\dagger_+(-x)\psi_+(x') \rangle \langle \psi_+(x) \psi^\dagger_+(-x') \rangle +\langle \psi^\dagger_-(x) \psi_-(-x') \rangle \langle \psi_-(-x) \psi^\dagger_-(x') \rangle + \langle \psi^\dagger_-(x) \psi_-(x') \rangle \langle \psi_-(-x) \psi^\dagger_-(-x') \rangle \\
&+\langle \psi^\dagger_-(-x) \psi_-(-x') \rangle \langle \psi_-(x) \psi^\dagger_-(x') \rangle+ \langle \psi^\dagger_-(-x) \psi_-(x') \rangle \langle \psi_-(x) \psi^\dagger_-(-x') \rangle -\langle \psi^\dagger_+(x) \psi_+(-x') \rangle \langle \psi_-(x) \psi^\dagger_-(-x') \rangle \\
 & +\langle \psi^\dagger_+(x) \psi_+(x') \rangle \langle \psi_-(x) \psi^\dagger_-(x') \rangle +\langle \psi^\dagger_+(-x) \psi_+(-x') \rangle \langle \psi_-(-x) \psi^\dagger_-(-x') \rangle - \langle \psi^\dagger_+(-x) \psi_+(x') \rangle \langle \psi_-(-x) \psi^\dagger_-(x') \rangle\\
& - \langle \psi^\dagger_-(-x) \psi_-(x') \rangle \langle \psi_+(-x) \psi^\dagger_+(x')\rangle +\langle \psi^\dagger_-(-x) \psi_-(-x')\rangle \langle \psi_+(-x) \psi^\dagger_+(-x') \rangle + \langle \psi^\dagger_-(x) \psi_-(x')\rangle \langle \psi_+(x) \psi^\dagger_+(x') \rangle \\
& -\langle \psi^\dagger_-(x) \psi_-(-x') \rangle \langle \psi_+(x) \psi^\dagger_+(-x') \rangle + \langle \psi^\dagger_+(x) \psi_+(-x) \rangle \langle \psi^\dagger_-(x')\psi_-(-x') \rangle +\langle \psi^\dagger_+(x) \psi_+(-x) \rangle \langle \psi^\dagger_-(-x') \psi_-(x') \rangle\\
&+ \langle \psi^\dagger_+(-x) \psi_+(x) \rangle \langle \psi^\dagger_-(-x') \psi_-(x') \rangle+\langle \psi^\dagger_-(x) \psi_-(-x) \rangle \langle \psi^\dagger_+(x') \psi_+(-x') \rangle +\langle \psi^\dagger_-(x) \psi_-(-x) \rangle \langle \psi^\dagger_+(-x') \psi_+(x') \rangle\\
& +\langle \psi^\dagger_-(-x) \psi_-(x) \rangle \langle \psi^\dagger_+(x') \psi_+(-x') \rangle +\left.\langle \psi^\dagger_-(-x) \psi_-(x)\rangle \langle \psi^\dagger_+(-x') \psi_+(x') \rangle+\langle \psi^\dagger_+(-x) \psi_+(x) \rangle \langle \psi^\dagger_-(x') \psi_-(-x') \rangle \right].
\end{split}
\label{eq:mag_corr}
\eeq
\end{widetext}
Using Eq. (\ref{eq:psi_evolution}) and the mode expansion of the fermionic field we can compute the average values of the products of pairs of fermionic operators, which are given by
\begin{subequations}
\beq
\begin{split}
\langle \psi^\dagger_{+,-}(x) \psi_{+,-}(y) \rangle =& e^{\pm i m(t-\abs{x}) \Theta(\pm x)} e^{\mp i m(t-\abs{y}) \Theta(\pm y)}\\
 &\frac{1}{2 \pi \left(\alpha\mp i (x-y)\right)}
\end{split}
\eeq
\beq
\begin{split}
\langle \psi_{+,-}(x) \psi^\dagger_{+,-}(y) \rangle =& e^{\pm i m(t-\abs{y}) \Theta(\pm y)} e^{\mp i m(t-\abs{x}) \Theta(\pm x)}\\
& \frac{1}{2 \pi \left(\alpha\pm i (x-y)\right)}.
\end{split}
\eeq
\end{subequations}
With these two expression we can then compute all the terms of Eq. (\ref{eq:mag_corr}) and, after some algebra, get the expression (\ref{eq:correlations}).
\\

\section{Coefficients of the quartic protocol}
\label{sec:coefficients}

In this appendix we give the explicit expressions for the coefficients of the quartic protocols considered in the case of the global protocols in the bosons system and in the Ising chain. 

In all three cases we wrote the coefficients $\rho_n$ as
\begin{subequations}
\beq
\rho_4=e_1/4,
\eeq
\beq
\rho_3=-\frac{e_1}{3}(\frac{1}{3}+e_2+e_3),
\eeq
\beq
\rho_2=\frac{e_1}{2}(\frac{e_2}{3}+\frac{e_3}{3}+ e_2 e_3 ),
\eeq
\beq
\rho_1=- \frac{e_1 e_2 e_3}{3},
\eeq
\end{subequations}
Then, for the bosons we set
\begin{subequations}
\beq
e_1=18+\frac{486 m_0}{m_0-m_1},
\eeq
\beq
\begin{split}
e_{2/3}=&\frac{1}{1344 m_0-48 m_1} \Bigl(580 m_0-13 m_1 \\
& \pm \sqrt{261136 m_0^2-9704 m_0 m_1 +73 m_1^2} \Bigr).
\end{split}
\eeq
\end{subequations}
In the case of the Ising chain and protocols starting and ending in the same phase we set
\begin{subequations}
\beq
e_1= \frac{9 (56 g_0-2 g_1-27)}{g_0-g_1},
\eeq
\beq
\begin{split}
e_{2/3}=&\frac{1}{48(56 g_0-27-2 g_1)} \Bigl[1160 g_0-26 g_1-567\\
&\bigl(251505+1044544 g_0^2+4 g_1(4779+73 g_1)\\
&-16 g_0 (64071+2426 g_1) \bigr)^{1/2}\Bigr].
\end{split}
\eeq
\end{subequations}
Finally for the case of protocols starting and ending in different phases we set
\begin{subequations}
\beq
e_1=\frac{27 (8 g_0-2 g_1-3)}{2 (g_0-g_1)}
\eeq
\beq
\begin{split}
e_{2/3}=&\frac{1}{12(8 g_0-2 g_1-3)} \Bigl[14 g_0-8 g_1-3+ \bigl(873+4676 g_0^2\\
&+32 g_1(18+g_1) -4 g_0 (1017+304 g_1) \bigr]^{1/2}
\end{split}
\eeq
\end{subequations}
 
\bibliography{Nonequilibrium}

\end{document}